\documentclass[fleqn,usenatbib,useAMS]{mnras}

\usepackage{hyperref}
\hypersetup{%
    colorlinks=true, 
    citecolor=blue} 

\usepackage[utf8]{inputenc}
\usepackage{amsmath}
\usepackage{upgreek}
\usepackage{amssymb}
\usepackage{color}
\usepackage[dvipsnames]{xcolor}
\usepackage{graphicx}	% Including figure files
\usepackage{supertabular}
\usepackage{caption}
\title[QPOs from post-shock accretion column of polars]{Quasi-periodic oscillations from post-shock accretion column of polars}
\author[P. Bera \& D. Bhattacharya]{Prasanta Bera\thanks{E-mail:pbera@iucaa.in}, Dipankar Bhattacharya\\Inter University Centre for Astronomy and Astrophysics, Post Bag 4, Pune 411007, India.}
\begin{document}
\pagerange{\pageref{firstpage}--\pageref{lastpage}} \pubyear{0000}

\maketitle
\label{firstpage}

\begin{abstract}
A set of strongly magnetized accreting white dwarfs (polars) shows quasi-periodic oscillations (QPOs) with frequency about a Hz in their optical luminosity. These Hz-frequency QPOs are thought to be generated by intensity variations of the emitted radiation originating at the post-shock accretion column. Thermal instability in the post-shock region, triggered by efficient cooling process at the base, is believed to be the primary reason behind the temporal variability. Here we study the structure and the dynamical properties of the post-shock accretion column including the effects of bremsstrahlung and cyclotron radiation. We find that the presence of significant cyclotron emission in optical band reduces the overall variability of the post-shock region. In the case of a larger post-shock region above the stellar surface, the effects of stratification due to stellar gravity becomes important. An accretion column, influenced by the strong gravity, has a smaller variability as the strength of the thermal instability at the base of the column is reduced. On the other hand, the cool, dense plasma, accumulated just above the stellar surface, may enhance the post-shock variability due to the propagation of magnetic perturbations. These characteristics of the post-shock region are consistent with the observed properties of V834~Cen and in general with Cataclysmic Variable sources that exhibit QPO frequency of about a Hz.
\end{abstract}

\begin{keywords}
stars: cataclysmic variables --- stars: magnetic field --- accretion --- methods: numerical --- radiation mechanisms: non-thermal
\end{keywords}

\section{Introduction}

Cataclysmic variables (CVs) are a kind of binary stellar systems which show sharp variations in their luminosity. It is thought that in such a binary system, a white dwarf, as a primary, accretes matter from its companion star  and the variation in the observed intensity is due to changes in the accretion rate. The secondary star loses mass either from the loosely bound outer layers in its late evolutionary phase or by outflowing stellar wind in its main sequence phase. In general, the accreting matter forms a disk to transport angular momentum after crossing the $L_1$ Lagrange point of the binary system. The accreted material is hot and ionised. Hence the presence of strong external magnetic field influences the motion of this ionised material. For a magnetic white dwarf, the disk is truncated at a place where the pressure due to the white dwarf magnetic field dominates the disk pressure, and the accreting matter follows the magnetic field lines from the disk to the magnetic pole. As the matter approaches the white dwarf, it gains kinetic energy at the expense of gravitational potential and the inflow velocity increases. The maximum free fall velocity is achieved near the stellar surface and a standing shock is formed just before the matter comes to rest at the surface. Polars are magnetic CVs in which the magnetic field  of the white dwarf is so strong that it prevents the formation of the accretion disk altogether. The accreting matter directly follows the magnetic field lines from $L_1$ and moves towards the magnetic pole. The binary system, in this case, is strongly coupled with the white dwarf spin fully synchronised with the binary orbital period.

The magnetically channelled hot plasma of the post-shock region cools by radiation. The connection between the observed X-ray radiation and the emission from the post-shock accretion column of the white dwarf was first suggested by \cite{Hoshi1973}. The thermal bremsstrahlung process is considered as the dominating cooling process, in which electrons, influenced by the ionic potential, emit X-ray photons \citep{King+Lasota1979}. Another source of radiation in this magnetic post-shock region is the cyclotron process \citep{Fabian+1976, Lamb+Masters1979}: the electron radiates in the optical or ultra-violet energy band while gyrating around magnetic field lines. Strong polarisation of the observed optical emission is attributed to such cyclotron emission. The magnetic field strength at the surface of the white dwarf is measured via the spectral signature of Zeeman effect as well as cyclotron lines \citep{Visvanathan+Wickramasinghe1979,Barrett+Chanmugam1985}.

Theoretical studies by \cite{Langer+1981} \& \cite{Langer+1982} suggest the possibility of thermal instability affecting an accretion column that emits only due to bremsstrahlung. Later, optical observations identified 0.1-1 Hz quasi-periodic oscillations (QPOs) with 1-3\% rms amplitude in some of these systems e.g. AN UMa \citep{Middleditch1982}, V834 Cen \citep{Middleditch1982, Mason+1983, Larsson1985}, EF Eri \citep{Larsson1987}, VV Pup \citep{Larsson1989}, BL Hyi \citep{Middleditch+1997} (Table~\ref{qpo_sources}). It is believed that these QPOs are the observational signatures of thermal instability in the accretion column. The instability causes periodic variation in the height of post-shock column leading to the variation of luminosity. In this scenario, the X-ray emission generated in this post-shock region is also expected to exhibit similar QPOs. But no QPO has been detected in the X-ray observations of these polars \citep{Bonnet-Bidaud+2015, Beardmore+Osborne1997a}, even in observations simultaneous with those in which optical QPOs are present \citep{Imamura+2000}. The best upper limits on the rms amplitude of X-ray QPOs in 0.1--5~Hz range from XMM-Newton observations are around $\sim 10$\% \citep{Bonnet-Bidaud+2015}. A  different class of QPOs, with periods of order minutes (e.g. 5-6 minutes QPOs from IGR J14536-5522 \cite{Potter+2010}), is seen in a few other magnetic CVs. These are thought to arise in instabilities in the zone of coupling of the accreting matter to the stellar magnetic field, far away from the surface of the white dwarf.

The structure of the accretion column on the white dwarf surface has been a subject of investigation over several decades. \cite{Aizu1973} modelled the post-shock column structure with bremsstrahlung and derived the expected X-ray spectrum from it. The column structure for different cooling processes has been studied by \cite{Wu+1994}. The mass of the white dwarf could be estimated using the derived model spectrum from the post shock structure \citep{Wu+1995}. \cite{Cropper+1999} derived the mass of the white dwarf more accurately by taking into account the vertical profile of white dwarf gravity along the column. The column structure and the radiated spectrum for a two temperature fluid, where the electrons and ions do not have enough time to equilibrate, have been studied by \cite{Saxton+2005}.

The stability property of the accretion column is obtained by studying the dynamical evolution of perturbations in the post-shock region. The dynamical properties of the post-shock region in the linear regime show a stable or unstable nature of the oscillation modes \citep{Chevalier+Imamura1982, Mignone2005}. This linear evolution is valid only on a short time-scale;  complex interactions between modes generate a different behaviour at a later stage. \cite{Imamura+1984, Imamura1985} have studied the characteristics of different oscillation patterns in the radiative shock. Inclusion of the cyclotron emission in the optical band, in addition to the bremsstrahlung X-rays, in the cooling of the post-shock matter results in a reduction of the oscillation amplitude and an increase of the QPO frequency \citep{Chanmugam+1985, Imamura+1991, Wu+1992, Saxton+1998}. Two-temperature effects also change the stability characteristics of a few modes \citep{Imamura+1996, Saxton+Wu1999}.

Non-linear dynamical evolution of the oscillating column for various cooling functions has been studied in both one and two-dimensional geometry \citep{Strickland+Blondin1995, Mignone2005}. \cite{Mignone2005} studied the effects of different possible boundary conditions at the base of the accretion column. The effects of cyclotron emission in the dynamical evolution has been found to reduce the oscillation amplitude \cite{Busschaert+2015}.

As the accretion proceeds, the accreted matter begins to accumulate at the base of the accretion column. The vertically immobile accumulated plasma remains confined to the column and is stratified due to the gravitational force of white dwarf. This gravitationally stratified layer, stored at the base of the accretion column, tries to escape horizontally from near the base and hence bends the threaded magnetic field lines \citep{Hameury+1983}. The bending becomes substantial when the accumulated mass is high enough to trigger a ballooning mode instability \citep{Litwin+2001}. 
 
In this paper, we aim to study the effects of magnetic field on the structure of a radiating accretion column and its temporal behaviour. In the section~\ref{section_model_equation}, we describe the basic equations used to model the accretion column. Section~\ref{section_method} describes the method used to solve the basic equations. The results are briefly summarised in section~\ref{section_results}. After comparing the results with the observations in section~\ref{observations_comparison}, we conclude in section~\ref{section_conclusion}.

\begin{table*}
 \caption{Polars with observed optical QPOs in the frequency range about a Hz. \textbf{References:} (1) \protect\cite{Middleditch1982}, (2) \protect\cite{Bonnet-Bidaud+1996}, (3) \protect\cite{Schwope+1993}, (4) \protect\cite{Larsson1987}, (5) \protect\cite{Howell+2006b}, (6) \protect\cite{Larsson1989}, (7) \protect\cite{Bonnet-Bidaud+2015}, (8) \protect\cite{Middleditch+1997}, (9) \protect\cite{Wolff+1999}, (10) \protect\cite{Mouchet+2017}}
 \label{qpo_sources }
 \begin{supertabular}{c|c|c|c c| c|c c|c c}
  \hline
    \bf 	&amplitude		&frequency	&\multicolumn{2}{|c|}{polarisation}&Mass		&B	&		&\\
    \bf Sources	&rms (\%)		&(Hz)		&Linear(\%)	&circular(\%)	&(M$\odot$)	& ($10^2$ T)	&	&references	& \\
  \hline
	AN UMa	&	2.4	&	0.4-0.8	&	9	&	-30	&	1	&29-36	&	&(1), (2)	& \\
	V 834 Cen&	1.2	&	0.4-0.8	&	-	&	15	&	0.66	&23	&	&(1), (3), (10)	& \\
	EF Eri	&	1.3	&	0.3-0.9	&	$\sim9$	&	$\sim9$	&	0.6	&30-59	&	&(4), (1), (5)	& \\
	VV Puppis&	1.8	&	$\sim1$	&	15	&	+8 to +15&	0.73	&$>30$	&	&(6), (1), (7)	& \\
	BL Hyi	&	1.-4.	&	0.2-0.8	&	-	&	-	&	1	&12-23	&	&(8), (9)	& \\
	%IGRJ14536-5522&	$<1$(??)&	0.0028-0.0032	&	-	&	$\sim0$ to $\sim$-18	&	-	&-	&	&(10)	& \\
  \hline
 \end{supertabular}
 \label{qpo_sources}
\end{table*}

\section{Model equations} \label{section_model_equation}

 In polars the accretion flow is influenced by the strong magnetic field of the white dwarf. We are interested in the dynamics of the accretion flow near the white dwarf surface where the matter moves along the magnetic field with free fall velocity and creates a dynamical shock just before impacting on the stellar surface. In this condition, the accretion flow can be described via the following hydrodynamic equations,
\begin{align}
 \frac{\partial\rho}{\partial t} + \nabla\cdot(\rho\bf{v}) &= 0;\label{continuity}\\
 \rho\left(\frac{\partial\mathbf{v}}{\partial t}+(\mathbf{v}\cdot\nabla)\mathbf{v}\right) &= -\mathbf{\nabla} p -\rho\mathbf{\nabla} \Phi_g; \\
 \frac{\partial(E+\rho\Phi_g)}{\partial t}+\nabla\cdot(\mathbf{v}(E+p+\rho\Phi_g)) &=  -\rho\mathbf{v\cdot\nabla} \Phi_g - \Lambda. \label{energy_conservation}
\end{align}
Here, $\rho$, $\mathbf{v}$, $p$ are density, velocity and pressure of the plasma respectively. $\frac{\partial}{\partial t}$ and $\nabla$ represent the temporal differentiation and the spatial gradient operator respectively. $\Phi_g$ is the gravitational potential, $E=\frac{1}{2}\rho v^2+\rho\epsilon$ the total energy density and $\epsilon=\frac{1}{\rho}\frac{p}{\gamma-1}$ the internal energy per unit mass. The term $\Lambda$ represents the cooling function i.e. the energy loss term. The energy radiated from the post-shock region influences the dynamics of the flow.

 \subsubsection*{1d-structure : }
 The geometrical structure of the accretion column depends on the height of the shock and the cross-sectional area of the magnetic polar cap over which the accretion flow impacts the stellar surface. A compact star of radius $R_{wd}$ with an accretion disc truncated at the Alfv\'en radius $r_{A}$ has a polar cap radius $r_p\sim R_{wd}\sqrt{\frac{R_{wd}}{r_A}}$ in a dipolar field geometry. In the case of a polar, without any accretion disk, $r_A$ may be replaced by the distance of the $L_1$ point from the white dwarf. The length along the accretion column from the stellar surface is denoted by $x$ and the shock height $x_s$ can be estimated\footnote{A more precise value of $x_s$ is calculated using equation~\ref{eq_Xs}.} as \citep{Wu+1994},
 \begin{align}
  x_s \sim 3\times10^5\textrm{m}\left[\frac{\dot{m}}{10~\rm{kg~m^{-2}~s^{-1}}}\right]^{-1}\left[\frac{M_{wd}}{0.5~M_\odot}\right]^\frac{3}{2}\left[\frac{R_{wd}}{10^7\textrm{m}}\right]^{-\frac{3}{2}}
 \end{align}
 where $\dot{m}$ is the mass accretion rate per unit area.
 
 An accretion column with $x_s\ll r_p$ is geometrically one dimensional. On the other hand, if the shock height is comparable to the polar cap radius ($x_s\sim r_p$) but the magnetic field is strong enough to restrict the plasma motion along the field lines, the behaviour of the system may be considered physically as one-dimensional with  variation only along the column axis. For these one-dimensional geometries the equations (\ref{continuity}-\ref{energy_conservation}) can be represented as,
\begin{align}
 \frac{\partial\rho}{\partial t} + \rho\frac{\partial v}{\partial x} + v\frac{\partial\rho}{\partial x} &= 0;\label{1d_continuity}\\
 \frac{\partial v}{\partial t}+v\frac{\partial v}{\partial x} + \frac{1}{\rho}\frac{\partial p}{\partial x} &= -\frac{\partial\Phi_g}{\partial x}; \label{1d_momentum_conservation}\\
 \frac{\partial p}{\partial t}+\gamma p\frac{\partial v}{\partial x} + v\frac{\partial p}{\partial x} &= -(\gamma-1)\Lambda. \label{1d_energy_conservation}
\end{align}
The matter contained within the accretion column is not significant compared to the total mass and hence we can neglect self-gravity of the accreted material and the gravitational potential $\Phi_g$ can be assumed to be $\frac{-GM_{wd}}{R_{wd}+x}$.
The steady accretion column structure can be obtained from equations~\ref{1d_continuity}-\ref{1d_energy_conservation} by dropping the time derivatives:
\begin{align}
 \rho\frac{\partial v}{\partial x} + v\frac{\partial\rho}{\partial x} &= 0;\label{eq1d_continuity}\\
 v\frac{\partial v}{\partial x} + \frac{1}{\rho}\frac{\partial p}{\partial x} &= -\frac{\partial\Phi_g}{\partial x}; \label{eq1d_momentum_conservation}\\
 \gamma p\frac{\partial v}{\partial x} + v\frac{\partial p}{\partial x} &= -(\gamma-1)\Lambda. \label{eq1d_energy_conservation}
\end{align}
 
 Here in the one-dimensional geometry we consider the flow of ionized plasma towards the white dwarf surface with free-fall velocity $v_{in}$, density $\rho_{in}$ and pressure $p_{in}$ just before the shock. The variables ($\rho_s, v_s, p_s$) just after the shock are obtained from the Rankine-Hugoniot jump conditions: 
 \begin{align}
  \frac{\rho_{s}}{\rho_{in}} &= \frac{v_{in}}{v_{s}};\\
  \frac{v_s}{v_{in}} &= \frac{\gamma-1}{\gamma+1}+\frac{2\gamma}{\gamma+1}\frac{p_{in}}{\rho_{in}v_{in}^2};\\
  p_s &= \frac{2\gamma}{\gamma+1}\rho_{in}v_{in}^2-\frac{\gamma-1}{\gamma+1}p_{in}.
 \end{align}
 For a shock structure, in general, the variable $p_{in}$ is represented by the Mach number ($\mathcal{M}$), which is related to the flow variables as $\mathcal{M}^2=\frac{\rho_{in}v_{in}^2}{\gamma p_{in}}$. The temperature at the shock can be calculated from pressure and density, $T_s=\frac{\mu m_B p_s}{\rho_s k_B}$, where, $\mu$, $m_B$ and $k_B$ are the mean baryon weight, the baryon mass and the Boltzmann constant respectively.
 %\textcolor{red}{shock jump condition}
 \subsubsection*{Bremsstrahlung and cyclotron cooling :}
 Thermal bremsstrahlung from electrons in the hot ionised post-shock plasma is an effective cooling mechanism as the resulting X-rays encounter very little optical depth and are radiated freely \citep{Frank+2002}.  The spectral emissivity of the plasma with temperature T and density $\rho$ can be expressed as \citep{Longair2011, Rybicki+Lightman1979}:
 \begin{align}\label{eq_brem_nu}
  \varepsilon_\nu &= \left(\frac{2\pi}{3m_e k_B T}\right)^\frac{1}{2}\frac{2^5\pi  e^6}{(4\pi\epsilon_0)^3 3m_e c^3 m_B^2}\rho^2 \nonumber\\
	     &~~~~~\times g_B(\nu, T)\exp\left(-\frac{h\nu}{k_B T}\right).
 \end{align}
Now, assuming the \textit{Gaunt factor}~$g_B(\nu, T)$ to be independent of frequency ($\nu$), the cooling rate due to thermal bremsstrahlung is obtained by integrating the emissivity~($\varepsilon_\nu$) over the whole spectrum as given below,
\begin{align}\label{eq_Lambda_brem}
 \Lambda_{brem} &= \left(\frac{2\pi k_B}{3m_e}\right)^\frac{1}{2}\frac{2^5\pi  e^6}{(4\pi\epsilon_0)^3 3h m_e c^3 m_B^2}g_B\rho^2 T^{1/2} \nonumber\\
		&= \Lambda_0\rho^2 T^{1/2} \nonumber\\
		&\simeq 1.4\times10^8 \left(\frac{n_e}{10^{22}~\rm{m^{-3}}}\right)^2\left(\frac{T}{10^8~\rm{K}}\right)^{1/2}.
\end{align}
Here, assuming the gaunt factor $g_B\simeq1$, the estimated value of the constant is $\Lambda_0\sim5.1\times10^{13} ~\rm{J~m^3~kg^{-2}~K^{-1/2}~s^{-1}}$ and $n_e$ represents the electron number density. %\textcolor{red}{optically thin?}

 In the presence of the strong magnetic field in the accretion column, the emission due to cyclotron radiation is also an important process. In this process, the electrons emit electromagnetic radiation during their gyrating motion with angular frequency $\omega_c~=~\frac{eB}{m_e}$ about the magnetic field lines. The cyclotron emission at frequencies $\omega<\omega^*$ is self-absorbed and produces a Rayleigh-Jeans spectrum \citep{Masters+1977}. The cyclotron luminosity $L_{cyc}$ from a region with cross-sectional area $A$ and thickness $\textrm{x}$ can be expressed as:
 \begin{align}
  L_{cycl} = \frac{Ak_BT}{12\pi^2c^2}[\omega^*(\textrm{x})]^3
 \end{align}
The value of $\omega^*$ is almost linearly dependent on the cyclotron frequency according to the following relation:
\begin{align}
 \omega^* \approx 9.87~\left(\frac{\omega_p^2 \textrm{x}/\omega_c c}{10^7}\right)^{0.05}~\left(\frac{T}{10^8~\rm{K}}\right)^{1/2}~\omega_c
\end{align}
where $\omega_p~\left(=\sqrt{\frac{n_e e^2}{m_e\epsilon_0}},~ \epsilon_0\rm{~being~the~free~space~permittivity}\right)$ is the plasma frequency \citep{Chanmugam+Wagner1979, Wada+1980}. The cyclotron cooling rate per unit volume is given by \citep{Langer+1982}\footnote{in cgs unit $\Lambda_{cycl} = 5.7\times10^6\rm{erg~cm^{-3}~s^{-1}}\left(\frac{\rho}{4\times10^{-8}~\rm{g cm^{-3}}}\right)^\frac{3}{20}\\ 
~~~~~~~~~~~~~~~~~~~~\times\left(\frac{\textrm{x}^2}{10^{15}~\rm{cm^2}}\right)^\frac{-17}{40}\left(\frac{T}{10^8~\rm{K}}\right)^\frac{5}{2}\left(\frac{B}{10~\rm{MG}}\right)^\frac{57}{20}$}
\begin{align}\label{eq_Lambda_cycl}
 \Lambda_{cycl} &= \frac{1}{A}\frac{dL_{cycl}}{dx} = \frac{k_BT}{80\pi^2c^2}\frac{{\omega^*}^3}{\textrm{x}} \nonumber\\
		&\simeq 1.3\times10^6 \rm{J~m^{-3}~s^{-1}}\left(\frac{n_e}{10^{22}~\rm{m^{-3}}}\right)^{3/20}\nonumber\\
		&~~~\times\left(\frac{\textrm{x}}{10^5~\rm{m}}\right)^{-17/20}\left(\frac{T}{10^8~\rm{K}}\right)^{5/2}\left(\frac{B}{10^3~\rm{T}}\right)^{57/20}\nonumber\\
		&\simeq 2.6\times10^{-14}\rho^{3/20}T^{5/2}\nonumber\\
		&~~~\times\left(\frac{\rm{x}^2}{10^{11}~\rm{m}^2}\right)^{-\frac{17}{40}}\left(\frac{B}{10^3~\rm{T}}\right)^{57/20} ~\rm{J~m^{-3}~s^{-1}}.
\end{align}
 Here $\textrm{x}$ is the characteristic geometrical length scale of the region which emits due to the cyclotron process. We choose this length scale to be either the shock height or the polar-cap radius whichever is smaller.
 
 Hence, the cooling term ($\Lambda$) in the energy conservation equation \ref{energy_conservation} is the resultant effect of bremsstrahlung and cyclotron radiation \citep{Saxton+1998, Busschaert+2015}. $$\Lambda = \Lambda_{brem}+\Lambda_{cycl}=\Lambda_{brem}\left[1+\epsilon_s\left(\frac{p}{p_s}\right)^{\beta_1}\left(\frac{\rho}{\rho_s}\right)^{\beta_2}\right].$$ 
 Here, $\Lambda_{brem}$ and $\Lambda_{cycl}$ are given in equation~\ref{eq_Lambda_brem} \& \ref{eq_Lambda_cycl} and these terms can be expressed as a function of $\rho$, $p$ as below:
\begin{align}
 \Lambda_{brem} \equiv \Lambda_0\rho^2T^{1/2} = \Lambda_1\rho^{3/2}p^{1/2},
\end{align}
\begin{align}
 \Lambda_{cycl} \equiv \Lambda_{0,c}\rho^{3/20}T^{5/2} = \Lambda_{1,c}\left(\frac{\rho}{\rho_s}\right)^{-\frac{47}{20}}\left(\frac{p}{p_s}\right)^{\frac{5}{2}}.
\end{align}
The value of the exponents are $\beta_1 = 2; \beta_2=-77/20$. The dimensionless parameter $\epsilon_s \left(= \frac{\Lambda_{1,c}}{\Lambda_1\rho_s^{3/2}p_s^{1/2}}\right)$ indicates the relative contribution of cyclotron emission compared to bremsstrahlung at the shock front.

% $\Lambda_0\sim 5\times10^{20} \rm{erg}~\rm{cm}^3 g^{-2} K^{-1/2}s^{-1}$
% \begin{align}
% \Lambda_{0,c} = 1.2\times10^8 \rm{erg}~cm^{-3} s^{-1}\left(\frac{S}{10^{15}~ cm^2}\right)^{-\frac{17}{40}}\nonumber\\ 
% \left(\frac{B}{10~ \rm{MG}}\right)^{\frac{57}{20}}\left(\frac{\rho}{4\times 10^{-8}~\rm{g~cm}^{-3}}\right)^{\frac{3}{20}}\left(\frac{T}{10^8~ K}\right)^{\frac{5}{2}}
% \end{align}
 \subsubsection*{Time scales :}
 In a dynamical evolution, the relative importance of different processes may be gauged from their characteristic time scales. The cooling time scale can be estimated from the ratio of the internal energy density and the cooling rate per unit volume at the shock front. For example, the bremsstrahlung cooling time $t_{brem}=\frac{\rho\epsilon}{\Lambda_{brem}}=\frac{p^{1/2}}{(\gamma-1)\Lambda_0\rho^{3/2}\left(\frac{\mu m_B}{k_B}\right)^{1/2}}$.

Now from the Rankine-Hugoniot condition for a strong shock ($\mathcal{M}\rightarrow\infty$) we obtain, $\rho_s=4\rho_{in}=4\frac{\dot{m}}{v_{in}}$, $p_s=\frac{3}{4}\rho_{in}v_{in}^2$ and $T_s=\frac{3}{16}\frac{\mu m_B}{k_B}v_{in}^2$. Assuming the fully ionized state of the plasma, i.e. $\mu=0.5$, the cooling time scales due to bremsstrahlung and cyclotron processes can be expressed as mentioned below:

\begin{align}
t_{brem}\simeq&~4\times10^{-2}\textrm{~s~}\left(\frac{\dot{m}}{10~\rm{kg~m^{-2}~s^{-1}}}\right)^{-1}\left(\frac{v_{in}}{10^6\rm{m~s^{-1}}}\right)^{2};\\
t_{cycl}\simeq &~4\times10^3\textrm{~s~}\left(\frac{\dot{m}}{10~\rm{kg~m^{-2}~s^{-1}}}\right)^{17/20}\left(\frac{v_{in}}{10^6\rm{~m~s^{-1}}}\right)^{-77/20}\nonumber\\
&\times\left(\frac{\textrm{x}^2}{10^{11}~\rm{m^2}}\right)^{17/40}\left(\frac{B}{10^3\rm{~T}}\right)^{-57/20}.
\end{align}
The dimensionless ratio $\epsilon_s$ is expressed as \citep{Chanmugam1995, Wu+1992, Wu+1994}\footnote{$\epsilon_s =\frac{t_{brem}}{t_{cycl}}\sim 9\times10^{-3}\left(\frac{n_e}{10^{22}~\rm{m^{-3}}}\right)^{-1.85}\\ 
~~~~~~~~~~~~~~~~~~~~\times\left(\frac{\textrm{x}}{10^5~\rm{m}}\right)^{-17/20}\left(\frac{T}{10^8~\rm{K}}\right)^{2}\left(\frac{B}{10^3~\rm{T}}\right)^{2.85}$},
\begin{align}%\label{eq_define_epsilon_s}
 \epsilon_s&=\frac{t_{brem}}{t_{cycl}}\sim10^{-5}\left(\frac{\textrm{x}^2}{10^{11}\rm{m^2}}\right)^{-17/40}\nonumber\\
 &\times\left(\frac{B}{10^3\rm{~T}}\right)^\frac{57}{20}\left(\frac{\dot{m}}{10\rm{~kg~m^{-2}~s^{-1}}}\right)^\frac{-37}{20}\left(\frac{v_{in}}{10^6\rm{~m~s^{-1}}}\right)^\frac{117}{20}. \label{expression_epsilon_s}
\end{align}

In all the expressions above we have assumed that the plasma components are in local thermal equilibrium. The bremsstrahlung and the cyclotron emission come only from the electrons. If the plasma species, i.e. electrons and ions, have sufficient time for multiple collisions between themselves, then the entire plasma can be described by a single temperature. For a single temperature $T$ at the shock front, the electron-ion collision time ($t_{ei}$) is given by \citep{Dougal+Goldstein1958, Spitzer1962}
\begin{align}
 t_{ei} &= \frac{3\sqrt{2}\epsilon_0^2 m_Bm_e^{-1/2}(\pi k_BT)^{3/2}}{e^4n_i\ln\Lambda_C}\nonumber\\
 ~&\sim2\times10^{-5}\textrm{~s~}\left(\frac{\ln\Lambda_C}{20}\right)^{-1}\left(\frac{\dot{m}}{10\rm{~kg~m^{-2}~s^{-1}}}\right)^{-1}\nonumber\\
 &~~~~~~~\times\left(\frac{v_{in}}{10^6\rm{~m~s^{-1}}}\right)^4.
\end{align}
Here the Coulomb logarithm $\ln\Lambda_C$ depends on electron density ($n_e$) and temperature ($T$). As long as this collision time is shorter than the most efficient cooling process (i.e. $t_{ei}<$ min($t_{brem}$, $t_{cycl}$)), a single temperature description is valid for the column structure. The two temperature effect becomes more important for white dwarfs with mass more than about 1~M$_\odot$ \citep{Imamura+1987}.

 \subsubsection*{Matter accumulation at the polar cap :}
 
 The matter inside a white dwarf is highly dense where the electrons become Fermi degenerate and the ions form a crystalline structure. Accreting matter settling at the surface is hot and has a lower density, so Fermi degenerate condition may not be achieved. Observations of magnetic CVs indicate that thermal blackbody emission from the white dwarf surface have temperatures $\sim 10^5-10^6$~K, equivalent to an energy $\sim 10-100$~eV (e.g. \cite{Schwarz+1998, deMartino+2004, Landi+2009, Worpel+Schwope2015}). The physical state of matter at the white dwarf surface can be identified from the value of the electrostatic coupling parameter $\Gamma=\frac{Z^2e^2}{4\pi\epsilon_0~k_BT}\left(\frac{4\pi n}{3}\right)^\frac{1}{3}\simeq 1.1\left(\frac{Z^2}{A^\frac{1}{3}}\right)\left(\frac{\rho}{10^{11}~\rm{kg~m^{-3}}}\right)^\frac{1}{3}\left(\frac{T}{10^8~K}\right)^{-1}$ \citep{Brown+Bildsten1998}. Hence for the temperature range $10^5-10^6$~K, matter with density $1-10^6~\rm{kg~m^{-3}}$ behaves as a non-degenerate system (as the Fermi energy $E_F\lesssim k_BT$) with  $\Gamma \sim 0.02-20$, i.e. the matter stays in a liquid or a gas phase. So for the accumulated matter, we can assume that the overall structure is described as an isothermal state whereas the non-thermal perturbed region behaves as an ideal gas with adiabatic constant $\gamma=\frac{5}{3}$. 

 \section{Method} \label{section_method}
 In this section we describe the method to solve the accretion column structure and to find its dynamical properties. The method used for calculation of matter accumulation is described in Appendix~\ref{GS_solution_methos}.
 \subsubsection*{Steady solution :}
 To find the steady solution we need to solve equations~\ref{eq1d_continuity}-\ref{eq1d_energy_conservation} simultaneously with proper boundary conditions. We assume that the plasma moves at a velocity $v_{in}$ with  density $\rho_{in}$ and Mach number $\mathcal M$ just before the shock, formed at a height $x_s$ from the white dwarf surface. It is also assumed that the plasma in the post-shock region behaves as an ideal adiabatic gas with index $\gamma=\frac{5}{3}$. The physical variables and the fluid equations are transformed into a dimensionless form using the characteristics values $\rho_{in}, |v_{in}|$ and $x_s$. Hence, $x=x_s\xi$, $\rho=\rho_{in}\rho_0,~v = |v_{in}|v_0,~v_{in}=-|v_{in}|,~p=\rho_{in}|v_{in}|^2p_0$ and so on. The equations \ref{eq1d_continuity}~\&~\ref{eq1d_momentum_conservation} provide
 \begin{align}
  \rho_0 v_0 &= -1; \label{eq1d_steady_accetion_rate}\\
  \mathrm{and,}~~ -v_0+p_0 &= \kappa \label{eq1d_m}.
 \end{align}
 For a constant $\Phi_g$, the value of $\kappa$ is $1+\frac{1}{\gamma\mathcal{M}^2}=m~~\rm{(say)}$ but in general, $\kappa$ satisfies the following relation
 \begin{align}
  \frac{d\kappa}{d\xi}=-\rho_0(\xi)\frac{d(\Phi_g/|v_{in}|^2)}{d\xi}.
 \end{align}
 To evaluate $\kappa$, we need the density distribution $\rho_0(\xi)$ but we do not know this distribution before solving the problem. To simplify the problem we initially consider an uniform gravitational potential (i.e. $\Phi_g$ is a constant).
 Now, with the help of equation~\ref{eq1d_energy_conservation} the relation between the velocity and the column distance $\xi$ (=$\frac{x}{x_s}$ i.e. distance normalized with respect to the shock height) can be expressed as \citep{Wu+1994, Mignone2005},
 \begin{align}
  \xi(v_0) &= \frac{f(v_0)}{f(v_s)}; \label{xi_solution}\\
  \mathrm{where,}~&f(v_0)=\int_0^{v_0}(-y)^{2-\frac{1}{2}}\nonumber\\
  \times&\frac{y+\gamma(\kappa+y)-y\frac{d\kappa}{dy}}{(\kappa+y)^\frac{1}{2}\left[1+\epsilon_s\left(\frac{\kappa+y}{p_s}\right)^{\beta_1}\left(\frac{-1}{y\rho_s}\right)^{\beta_2}\right]}dy.
 \end{align}
 Here $v_0=0$ at $\xi=0$ and $v_0=v_s$ at $\xi = 1$. For a strong shock the value of $v_s$ is $\frac{1}{4}$. Using equation~\ref{xi_solution} one can get the steady velocity~($v_0$) at a distance~$\xi$. The other variables can be derived from this relation and equations~\ref{eq1d_steady_accetion_rate}~\&~\ref{eq1d_m}. The accretion shock height $x_s$ is related to the normalization constants by the following relation
 \begin{align}
  x_s\Lambda_1\rho_{in}|v_{in}|^{-2} = -\frac{f(v_s)}{\gamma-1}.\label{eq_Xs}
 \end{align}

 If the height of the post-shock region is very small compared to the white dwarf radius ($x_s\ll R_{wd}$), the effects of the gravitational term on the accretion column structure are not significant. To include the effects of the gravitational force due to the white dwarf on the equilibrium structure of the post shock region (i.e. $\Phi_g$ has a spatial dependency), we calculate the vertical profile iteratively, starting from the solution obtained using equation~\ref{xi_solution} \citep{Cropper+1999}. The value of $\kappa$ is derived from the solution of the previous iteration using the following relation 
 \begin{align}
  \kappa(\xi)=m-\int_{\xi=1}^{\xi}\rho_0(\xi)\frac{d(\Phi_g/|v_{in}|^2)}{d\xi}d\xi.
 \end{align}
 The iteration is terminated when the relative change of the variables becomes smaller than a predefined cutoff. For this one-dimensional accretion column geometry, the gravitational acceleration is expressed as $-\nabla\Phi_g=g_{norm}(1+\xi x_s/R_{wd})^{-2}x_s^{-1}v_{in}^2$. Here, $g_{norm}$ measures the strength of the gravitational acceleration in the normalized units at the stellar surface. For a white dwarf, the value of the stellar mass decides the inflow velocity which is nearly the free-fall velocity and hence the value of $g_{norm}$ is dependent on the shock height $x_s$.
 %\textcolor{red}{normalization constant ???}
 \subsubsection*{Linear evolution :} \label{method_linear}
 The linearized equations are obtained from the equations~\ref{1d_continuity}-\ref{1d_energy_conservation} by expressing the variables as a combination of the steady value and a perturbed quantity. After the perturbation, the shock position is given by
 \begin{align}
  x_s=x_{s_0}\left(1+\lambda_{x_s}\epsilon e^{\omega t}\right).
 \end{align}
  Here, $\lambda_{x_s}$ \& $\omega$ are complex variables, $\epsilon = 0$ represents the unperturbed condition and $\epsilon = 1$ gives the perturbed values. The normalized length scale $\xi$ along the shock is expressed as
 \begin{align}
  \xi = \frac{x}{x_s}\sim\frac{x}{x_{s_0}}\left(1-\lambda_{x_s}\epsilon e^{\omega t}\right).
 \end{align}

 For a perturbed shock front velocity amplitude $v_{s_1}$, the position displacement parameter $\lambda_{x_s}$ is described as $\lambda_{x_s}=\frac{v_{s_1}}{x_{s_0}\omega}=\frac{\lambda_{v_1}}{\delta}$, where $\delta=\frac{x_{s_0}\omega}{|v_{in}|}$. For a strong shock the boundary conditions of the perturbed variables at the shock are $\rho_1 = 0,~p_1=\frac{3}{2}\rho_{in}v_{in}v_{s_1},~v_1=\frac{3}{4}v_{s_1}$. The non-dimensional variables at a time $t$ can be expressed as \citep{Chevalier+Imamura1982, Imamura+1996, Saxton+1998}
 \begin{align}
  q(\xi, t) = q_0(\xi)[1+\lambda_q(\xi)\lambda_{v_1}\epsilon e^{\delta t}]
 \end{align}
 where $q\in\{\rho, v, p\}$, $q_0(\xi)$ represents the steady solution of the variables and $\lambda_q(\xi)$ is the complex parameter representing the perturbed component of the variables. In the comoving frame of the shock, the fluid equations~\ref{1d_continuity}-\ref{1d_energy_conservation} are converted to a linear form where the differentials take the form
 \begin{align}
  \frac{\partial}{\partial t}\rightarrow\frac{\partial}{\partial t}+\frac{\partial\xi}{\partial t}\frac{\partial}{\partial \xi},~~~ \frac{\partial}{\partial x}\rightarrow\frac{\partial\xi}{\partial x}\frac{\partial}{\partial \xi}.
 \end{align}
 The linearized equations for a constant gravitational potential are \citep{Mignone2005}:
 \begin{align}
  \frac{d\lambda_\rho}{dv_0}+\frac{d\lambda_v}{dv_0} &= -\frac{\xi}{v_0^2}-\frac{\lambda_\rho\delta}{v_0}\frac{d\xi}{dv_0},\label{eq_lin_continuity}\\
  v_0\frac{d\lambda_v}{dv_0}-p_0\frac{d\lambda_p}{dv_0} &= -\lambda_v\delta\frac{d\xi}{dv_0}+\frac{\xi}{v_0}+\lambda_p-2\lambda_v-\lambda_\rho,\label{eq_lin_momentum}\\
  v_0p_0\left(\gamma\frac{d\lambda_v}{dv_0}+\frac{d\lambda_p}{dv_0}\right) &= (v_0+\gamma p_0)\left[\frac{3}{2}\lambda_\rho-\frac{1}{2}\lambda_p -\lambda_v\right.\nonumber\\
  &+\mathcal{A}\left[\beta_1(\lambda_p-\lambda_{p_s})+\beta_2(\lambda_\rho-\lambda_{\rho_s})\right]\nonumber\\
  &\left.+\frac{1}{\delta}\right]-p_0\lambda_p\delta\frac{d\xi}{dv_0}+\xi\label{eq_lin_energy}.
 \end{align}
  Here $\mathcal{A} = \frac{\epsilon_s\left(\frac{p_0}{p_{0_s}}\right)^{\beta_1}\left(\frac{\rho_0}{\rho_{0_s}}\right)^{\beta_2}}{1+\epsilon_s\left(\frac{p_0}{p_{0_s}}\right)^{\beta_1}\left(\frac{\rho_0}{\rho_{0_s}}\right)^{\beta_2}}$.
 In the reduced perturbed variables the jump conditions at the shock front are:
 \begin{align}
  \lambda_\rho=0,~~~\lambda_v=-3,~~~\lambda_p=2.
 \end{align}
 The linearly perturbed equations~\ref{eq_lin_continuity}-\ref{eq_lin_energy} are integrated from the shock-front towards the shock base for a set of frequency components ($\delta_R$,  $\delta_I$ are real and imaginary components of frequency respectively). The mode frequencies ($\delta$) satisfy the stationary boundary condition at the stellar surface i.e. real and imaginary parts of the velocity perturbation ($\lambda_v$) identically vanish at $\xi=0$.
 \subsubsection*{Non-linear evolution :}
We solve the system of  dynamical evolution equations~\ref{1d_continuity}-\ref{1d_energy_conservation} using the magnetohydrodynamics code \textsc{pluto} \citep{Mignone+2007}. The initial condition is assumed as the equilibrium solution obtained from equation~\ref{xi_solution}. The upper boundary is assumed to have a steady inflow with a specified $\rho_{in}$ and $v_{in}$. The lower boundary is considered to be a thick absorbing layer. The downward flow velocity, density and pressure at this thick layer are kept constant at the equilibrium value corresponding to $x=10^{-3}x_s$. An outflow boundary condition is used at the end of this thick layer. In general,  about 2000 linearly spaced grid points are used to resolve the post-shock region, and the consistency of the result is checked with a higher resolution.
 
 Starting from the initial steady solution, the dynamical time evolution of this accretion shock structure is followed to study the effects of perturbations. Here we do not impose any external perturbation to the equilibrium solution, but the system gains it from numerical errors. The time evolution is performed in small time steps maintaining the CFL condition. The time integration of each step is done by the RK3 method while the Riemann problem is solved by the \textit{roe} solver.
 
 The cooling part is computed as a separate source term,
 \begin{align}
  \frac{dE}{dt} = - \Lambda(p,\rho). \label{eq_cooling_pluto}
 \end{align}
 In the post-shock region, the flow velocity gradually decreases towards the stellar surface and the internal energy is higher than in the pre-shock material. To include the effects of radiation from the plasma, it is assumed that the kinetic energy part remains unaltered while the pressure term of the total energy is modified due to cooling. In an integration time step, a modified pressure ($p$) is calculated from the cooling term which is dependent on $p$ and $\rho$. This time-integration step (from equation~\ref{eq_cooling_pluto}) is performed using RK4 method. The cooling effects are implemented only in the post-shock region by setting a lower threshold of temperature ($T_c$) such that the cooling is not effective at the pre-shock region and at the lower boundary layers.

 \section{Results} \label{section_results}

Here we briefly describe the results obtained from the steady and the dynamical evolution of the one-dimensional post-shock accretion column. The magnetic field geometry at the base of the accretion column and the criteria of magnetic instability are also mentioned for isothermal matter accumulation.

\subsection{1-d accretion column : Steady solution and X-ray spectrum}\label{subsec_eq_sol}
\begin{figure}
\centering
%\put(0,0){A.}
%\includegraphics[width=0.47\textwidth]{shock_equilibrium_alpha0_5.pdf}
\includegraphics[width=0.47\textwidth]{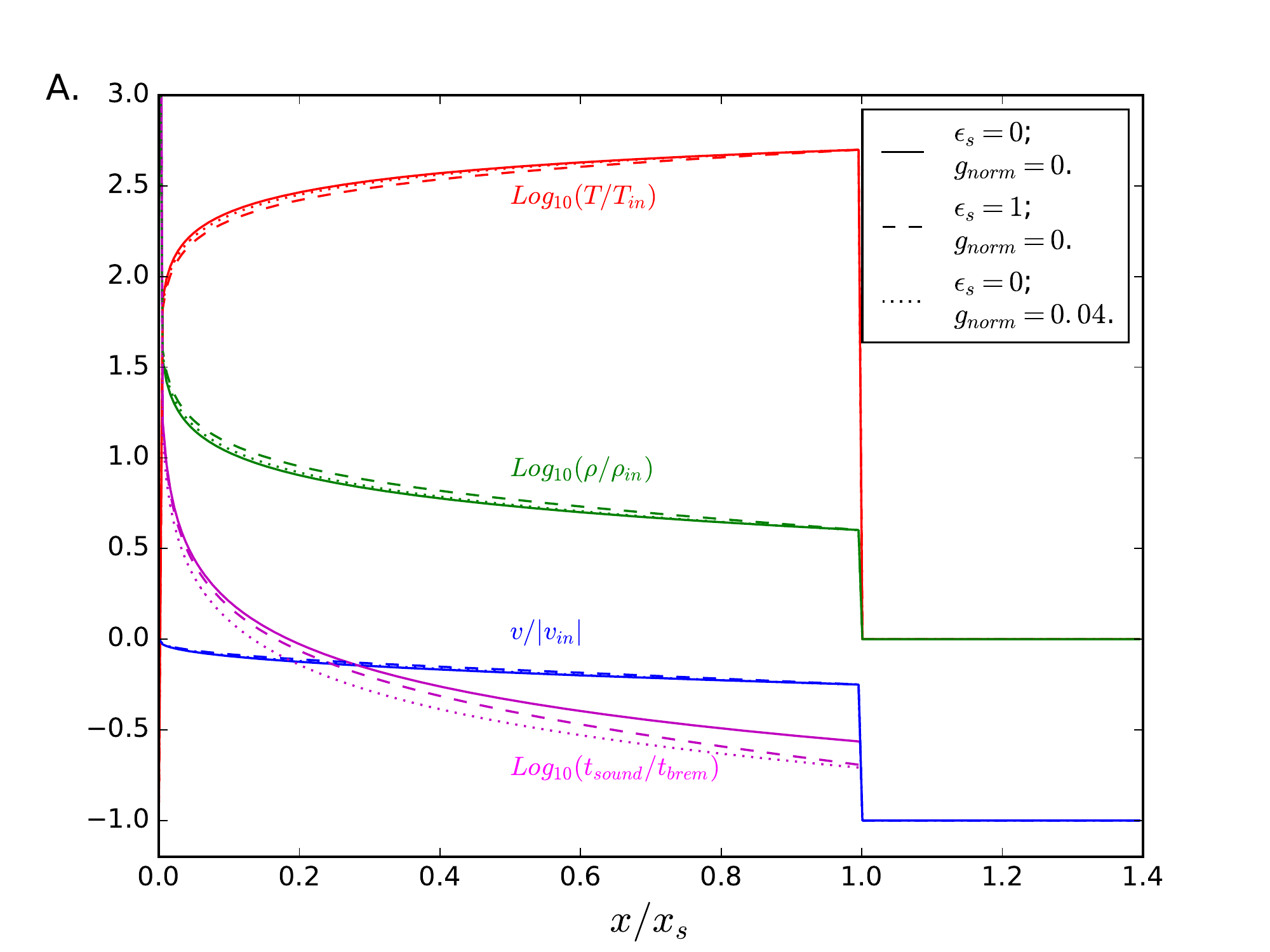}
\includegraphics[width=0.47\textwidth]{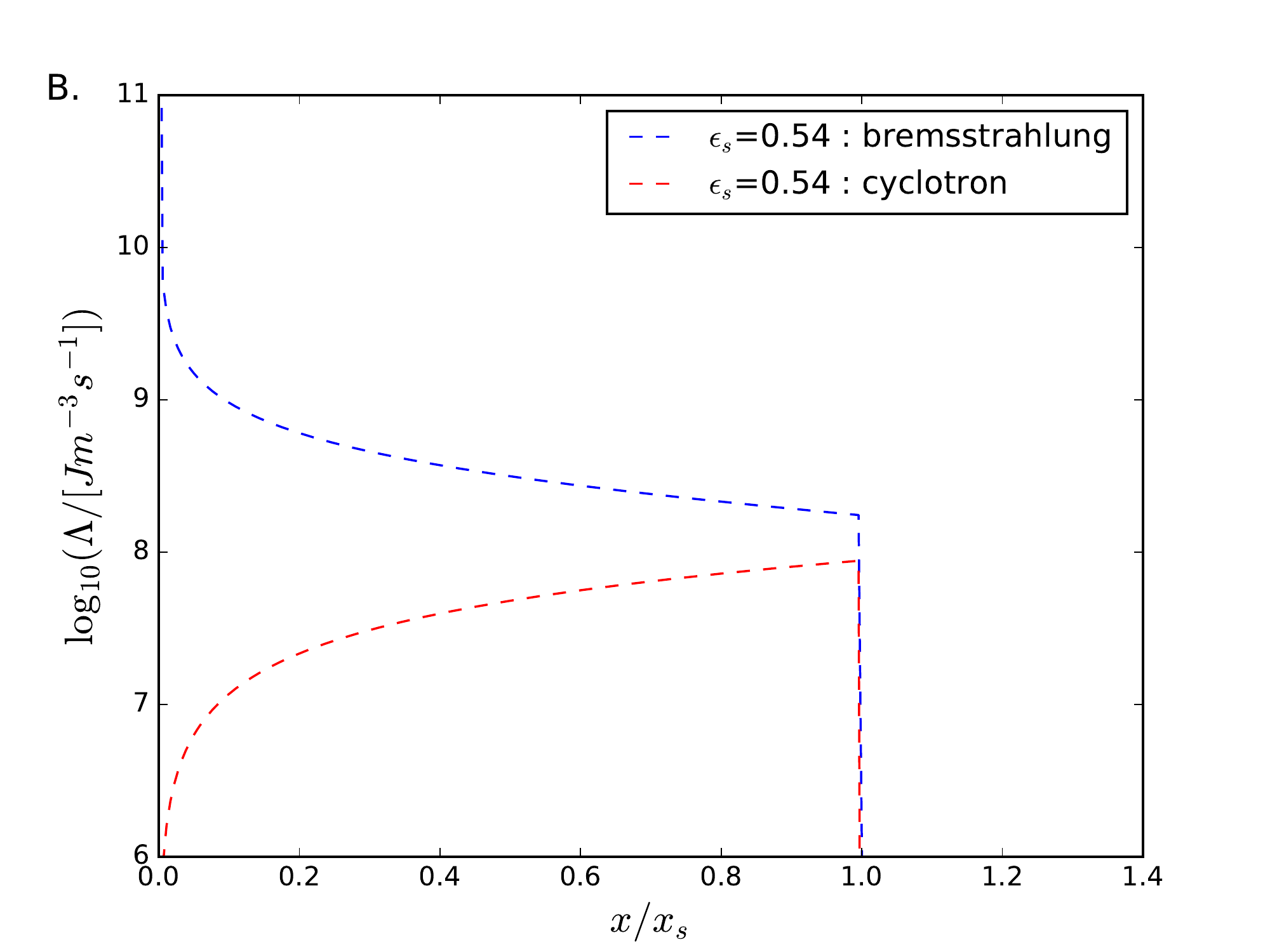}
\caption{\textbf{A.} The equilibrium solution of a steady shock with bremsstrahlung cooling in absence (continuous line) and presence (dotted line) of gravitational acceleration and with cyclotron cooling included  (dashed line). The plasma density ($\rho$), temperature ($T$), velocity ($v$) and the ratio of sound crossing time to the bremsstrahlung cooling time ($t_{sound}/t_{brem}$) are shown for the post shock region. \textbf{B.} Emission rate per unit volume in the two different processes--bremsstrahlung ($\Lambda_{brem}$) and cyclotron ($\Lambda_{cycl}$) from the post shock region. The bremsstrahlung emission is more efficient at the base of the column whereas the cyclotron process is more efficient near the shock front.} %\textcolor{red}{solution for $\epsilon_s\neq0$?}}
\label{shock_equilibrium}
\end{figure}

For an assumed steady accretion in a one-dimensional geometry, the density and velocity profiles of the matter in the shocked region are obtained by solving the equations~\ref{eq1d_continuity}-\ref{eq1d_energy_conservation}. The continuous curves of Fig.~\ref{shock_equilibrium} show the structure of the matter density, velocity and the temperature of the accretion shock, cooled by pure bremsstrahlung radiation. The effects of gravitational potential are neglected in this steady solution. The matter at $x > x_s$ approaches the white dwarf with a speed $v_{in}$, i.e. the free fall velocity, before forming a shock at $x = x_s$. For a strong shock, the matter velocity is about one-fourth of $v_{in}$ just behind the shock and the plasma comes to rest on the stellar surface ($x = 0$). In this steady accretion, the matter inflow rate is constant (equation~\ref{eq1d_steady_accetion_rate}) at every distance. Hence, in the accretion column structure, as the velocity tends to zero towards the surface the matter density correspondingly increases. If the shock height ($x_s$) from the stellar surface is not significant compared to the stellar radius ($x_s\ll R$), the assumption of an invariant gravitational potential in the shocked region remains valid. Hence the pressure profile is expected to be uniform unless modified by the effects of cooling. The matter is cool near the stellar surface as the bremsstrahlung X-ray emissivity depends more strongly on the density than the temperature ($\Lambda_{brem} \sim\rho^2T^{1/2}$).

The inclusion of the cyclotron process in the cooling rate modifies the equilibrium solution. The dashed lines in Fig.~\ref{shock_equilibrium}A show the profiles of the variables where the cooling due to cyclotron emission is included. Although the structure does not change significantly, the post-shock material is cooler and denser in comparison to the profiles obtained for bremsstrahlung alone. 

The emission rate from the post-shock region is obtained using the steady profile. Fig.~\ref{shock_equilibrium}B shows the emission rates due to bremsstrahlung and cyclotron processes from different parts of the accretion column. As already mentioned, bremsstrahlung is more strongly dependent on density compared to the temperature, and it dominates at the base of the column. On the other hand, the cyclotron process is strongly dependent on temperature, and hence it dominates near the shock front.

\begin{figure}
\centering
\includegraphics[width=0.47\textwidth]{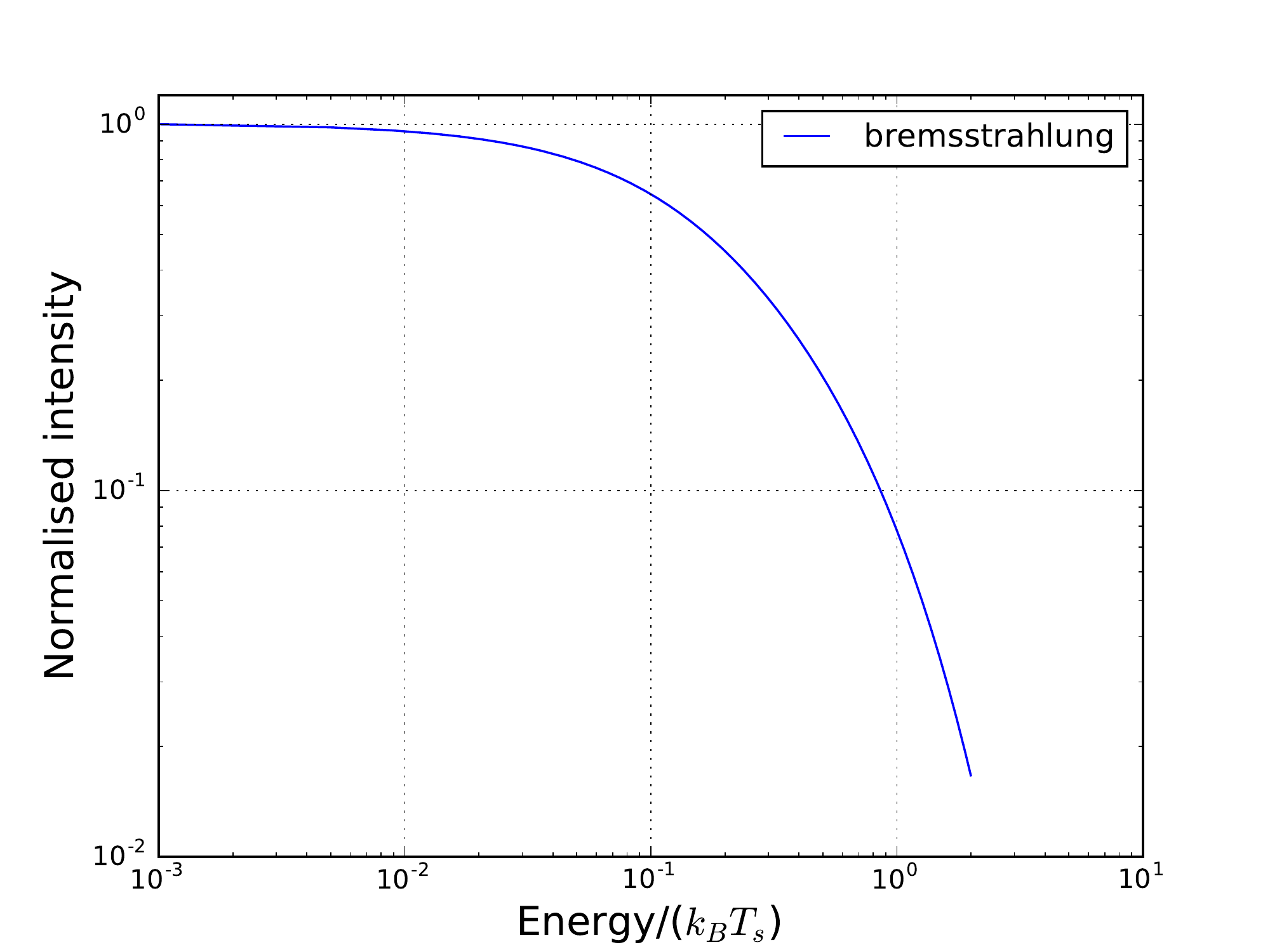}%{brem_spec.pdf}
\caption{The model bremsstrahlung spectrum from the accretion shock region. Here the Gaunt factor $g(\nu, T)$ is assumed to be a constant.}
\label{brem_spec}
\end{figure}

The energy dependent emissivity (equation~\ref{eq_brem_nu}) from different parts of the accretion column is summed to obtain the X-ray spectrum. Fig.~\ref{brem_spec} shows the bremsstrahlung spectrum due to the radiation from the accretion shock. The exponential high energy tail of the Maxwellian velocity distribution causes the exponential decay of the thermal bremsstrahlung spectrum at the high energy end. The resultant spectrum carries the signature of the high energy cutoff, corresponding to the shock temperature which is often used to estimate the mass of the white dwarf.

One of the accretion profiles shown in Fig.~\ref{shock_equilibrium}A is obtained by including the effects of gravitational acceleration within the column. The gravitational force of the white dwarf causes a stratification of the post shock structure and reduces the height of the post-shock region. The change in the post-shock profile also affect the dynamical characteristics, as described in the following sections. We use these steady solutions for studying the linear evolution of perturbations, and these are also considered as initial conditions for nonlinear dynamical evolution.

\subsection{1-d accretion column : Linear perturbation}

 \begin{table*}
 \caption{Linear oscillation modes of the post-shock accretion column for various values of the cyclotron cooling fraction ($\epsilon_s$).}
 \begin{supertabular}{c|c c|c c|c c|c c|c c}
  \hline
		&	\multicolumn{2}{|c|}{$\epsilon_s = 0$}	&	\multicolumn{2}{|c}{$\epsilon_s = 0.3$}	&\multicolumn{2}{|c}{$\epsilon_s = 1.0$}&\multicolumn{2}{|c}{$\epsilon_s = 10.0$}	&	\multicolumn{2}{|c}{$\epsilon_s = 100.0$}\\ \cline{2-11}
    \bf mode	&$\delta_R$	&$\delta_I$	&$\delta_R$	&$\delta_I$	&$\delta_R$	&$\delta_I$	&$\delta_R$&$\delta_I$	&$\delta_R$&$\delta_I$	\\
  \hline
	n=0	&	-0.0101	 &	0.3054	&	-0.0118&	0.3119	&-0.0160& 0.3202&-0.0514& 0.3166&-0.0973& 0.2334\\
	n=1	&	0.0476	 &	0.8889	&	0.0306	&	0.8754	&0.0075 & 0.8503&-0.0263& 0.7190&-0.0232& 0.5478\\
	n=2	&	0.0600	 &	1.5043	&	0.0429	&	1.4754	&0.0176 & 1.4275&-0.0653& 1.2137&-0.0960& 0.8326\\
	n=3	&	0.0848	 &	2.1077	&	0.0625	&	2.0605	&0.0301 & 1.9776&-0.0214& 1.6200&-0.0960& 1.2132\\
	n=4	&	0.0856	 &	2.7220	&	0.0645	&	2.6626	&0.0308 & 2.5609&-0.0852& 2.0845&-0.0247& 1.4510\\
	n=5	&	0.1041	 &	3.3285	&	0.0827	&	3.2520	&0.0519 & 3.1199&-0.0331& 2.5730&-0.1289& 1.7736\\
	n=6	&	0.1011	 &	3.9399	&	0.0784	&	3.8494	&0.0422 & 3.6925&-0.0492& 2.9710&-0.1058& 2.1513\\
	n=7	&	0.1164	 &	4.5426	&	0.0947	&	4.4382	&0.0617	& 4.2585&-0.0654& 3.4714&-0.0287& 2.3941\\
  \hline
 \end{supertabular}\label{linear_modes}
\end{table*}
The linearly perturbed equations are solved on the background steady solution with the specified boundary conditions to find the mode frequencies and the corresponding growth rates as mentioned in section~\ref{method_linear}. The first eight mode frequencies of the linear perturbation of a gravitationally non-stratified accretion column, radiating due to cyclotron as well as bremsstrahlung processes, are listed in Table~\ref{linear_modes}. The real parts ($\delta_R$) of the mode frequencies in the case of pure bremsstrahlung ($\epsilon_s = 0$) are positive except for the fundamental mode. Hence the linear perturbation study suggests a growth of instability for modes $n\geq 1$. As the influence of the magnetic field increases, the fraction of cooling due to the cyclotron process is also enhanced. For an accretion column with a higher $\epsilon_s$ value, i.e. with stronger cyclotron cooling effects, the mode characteristics remain similar but the mode frequencies for $n\geq 1$ decrease in comparison to their non-magnetic counterparts.  The frequency of the fundamental mode, however, increases. 

 \cite{Saxton+1998} calculate the real and imaginary parts of the eigenvalues for the first eight eigenmodes of an accretion shock with bremsstrahlung and cyclotron cooling. Here we find that the imaginary parts of the eigenvalues ($\delta_I$) match closely (within $\sim1\%$) with this solution. The real parts of the eigenvalues differ a little, but the relative trend of the variation and the sign match well.
 
\subsection{1-d accretion column : Non-linear evolution}
\label{subsec:perturb_nonlin}
\begin{figure*}
\centering
\includegraphics[width=0.99\textwidth]{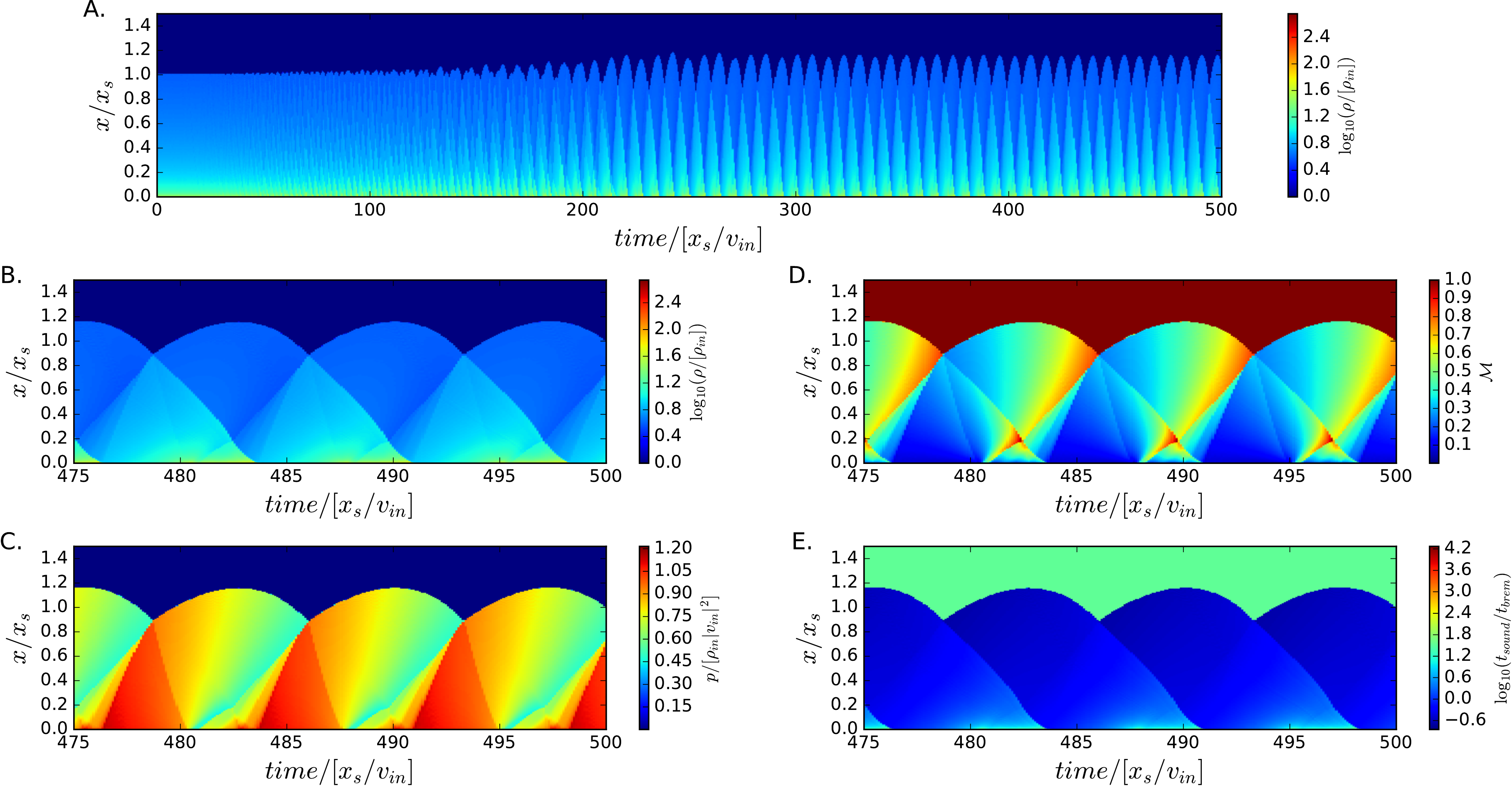}
\caption{\textbf{A.} The dynamical evolution of the 1-dimensional density profile in a radiative post-shock column where bremsstrahlung is the dominant cooling process. The evolution started from an equilibrium solution. The initially steady density profile gradually develops an oscillatory behaviour. At a later time the oscillation amplitude saturates. In the saturated phase, the time series plots of \textbf{B.} density, \textbf{C.} pressure, \textbf{D.} Mach number and \textbf{E.} the ratio between $t_{sound}$ to $t_{brem}$ show that the generation of a secondary shock ($t_{sound} > t_{brem}$) from the base and the propagation towards the primary shock front of the accretion column. }
\label{time_series_brem}
\end{figure*}

Non-linear dynamical evolution of the radiating accretion shock is computed using the magnetohydrodynamic code \textsc{pluto}. The equilibrium solution, obtained in \ref{subsec_eq_sol}, is used as the initial starting condition. An uniform inflow boundary, with constant mass inflow at free fall velocity with Mach number $\mathcal{M}=40$, is maintained at the top of the computational domain. The same amount of mass flux is removed at the base of the computational domain using a layer of thick absorbing medium. The time sequence of the density profile of the accretion column is shown in Fig.~\ref{time_series_brem} for the case of pure bremsstrahlung cooling in absence of gravitational stratification. At the early phase of the evolution, the shock position and the density profile retain their equilibrium values. But at a later time, the shock oscillates about its equilibrium position. The amplitude of the fluctuations becomes steady after a while. 

Time sequence plots of the density, pressure and Mach number demonstrate the movement of the discontinuities in density and pressure which are correlated with the oscillation phase. These characteristics of the flow have also been found in earlier studies by \cite{Strickland+Blondin1995, Sutherland+2003, Mignone2005, Busschaert+2015}. \cite{Busschaert+2015} attribute the propagation of the discontinuity to the presence of a secondary shock driven by the efficient bremsstrahlung cooling near the white dwarf surface. \cite{Falle1981} shows that a cooling rate $\Lambda\sim\rho^2T^\alpha$ generates a cooling catastrophe when $\alpha<\frac{3}{2}$, a condition which is satisfied for the bremsstrahlung radiation ($\alpha=\frac{1}{2}$). According to this theory, the secondary shock is formed when the local cooling timescale is shorter than the sound crossing time scale. Laboratory experiments in which a hot plasma flow collides with a rigid obstacle have demonstrated the presence of such secondary shocks \citep{Falize+2011, Cross+2016}. %\textcolor{red}{results of Plasma experiments}

Fig.~\ref{time_series_brem} clearly displays the secondary shock (with Mach number $\mathcal{M}\sim$ 1) starting near the base of the accretion column and propagating towards the primary shock front, causing a reduction of the shock height. After the collision of these shock fronts, a rarefaction wave propagates back to the base.  Increase of the shock height from the most compressed state causes a drop in matter pressure, driving another secondary shock from the base. This sequence of events recurs cyclically. To identify the origin of the oscillation we calculate the local cooling time due to bremsstrahlung $\left(t_{brem}=\frac{p}{(\gamma-1)}/(\Lambda_1\rho^\frac{3}{2}p^\frac{1}{2})\right)$ and compare with the sound crossing time scale ($t_{sound}\sim x_s/\sqrt{\frac{\gamma p}{\rho}}$). The profile of the ratio between the sound crossing time and the local bremsstrahlung cooling time indicates that the region with cooling catastrophe (i.e. $t_{sound}>t_{brem}$) appears close to the base of the post shock region (Fig.~\ref{shock_equilibrium}A, \ref{time_series_brem}E). The secondary shock originates as the cooling catastrophe begins and starts to propagate towards the primary shock. This satisfies Falle's criteria of the formation of the secondary shock in the post-shock accretion column.

\begin{figure*}
\centering
\includegraphics[width=0.99\textwidth]{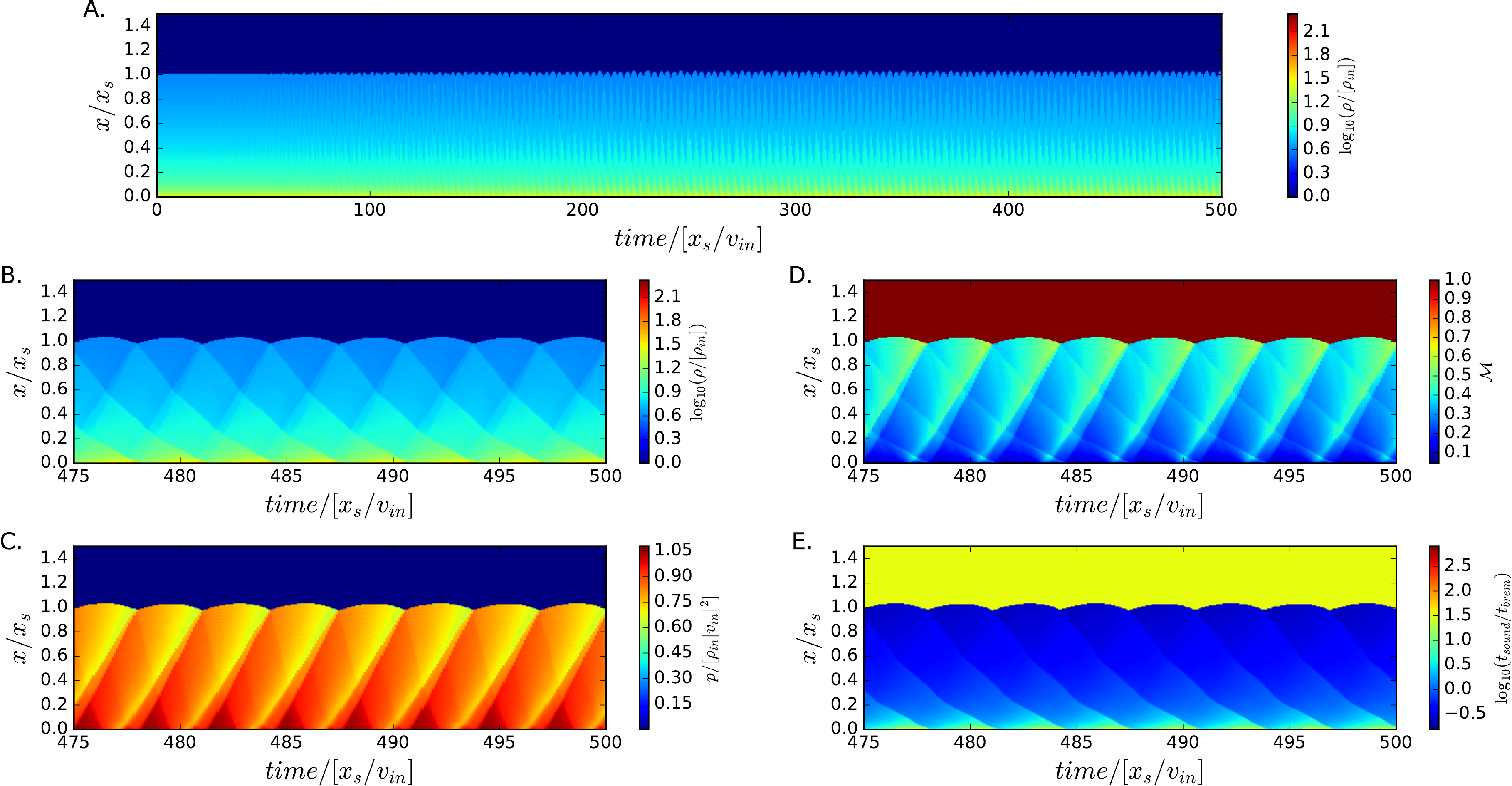}
\caption{Same as Fig.~\ref{time_series_brem} but including the effects of cyclotron emission ($\epsilon_s=0.7$). }
\label{time_series_combines}
\end{figure*}

Next, we include the effects of cyclotron emission in the dynamical study by evolving the equilibrium solution with non-vanishing $\epsilon_s$. The results are displayed in Fig.~\ref{time_series_combines}.  The density evolution exhibits the growth of the shock oscillation as in the case of pure bremsstrahlung, but the inclusion of cyclotron emission reduces the oscillation amplitude. This suppression of the oscillatory behaviour results from the cyclotron mechanism contributing additional efficient cooling near the primary shock, reducing the eventual strength of bremsstrahlung cooling catastrophe at the column base. We also find that compared to the pure bremsstrahlung case, higher order modes are more strongly excited in the presence of strong cyclotron cooling.  This is in accord with the results of the linear evolution, where a similar trend is observed (Table~\ref{linear_modes}). %\textcolor{red}{Estimation of change in oscillation time period due to cyclotron emission !!}

\begin{figure}
\centering
\includegraphics[width=0.47\textwidth]{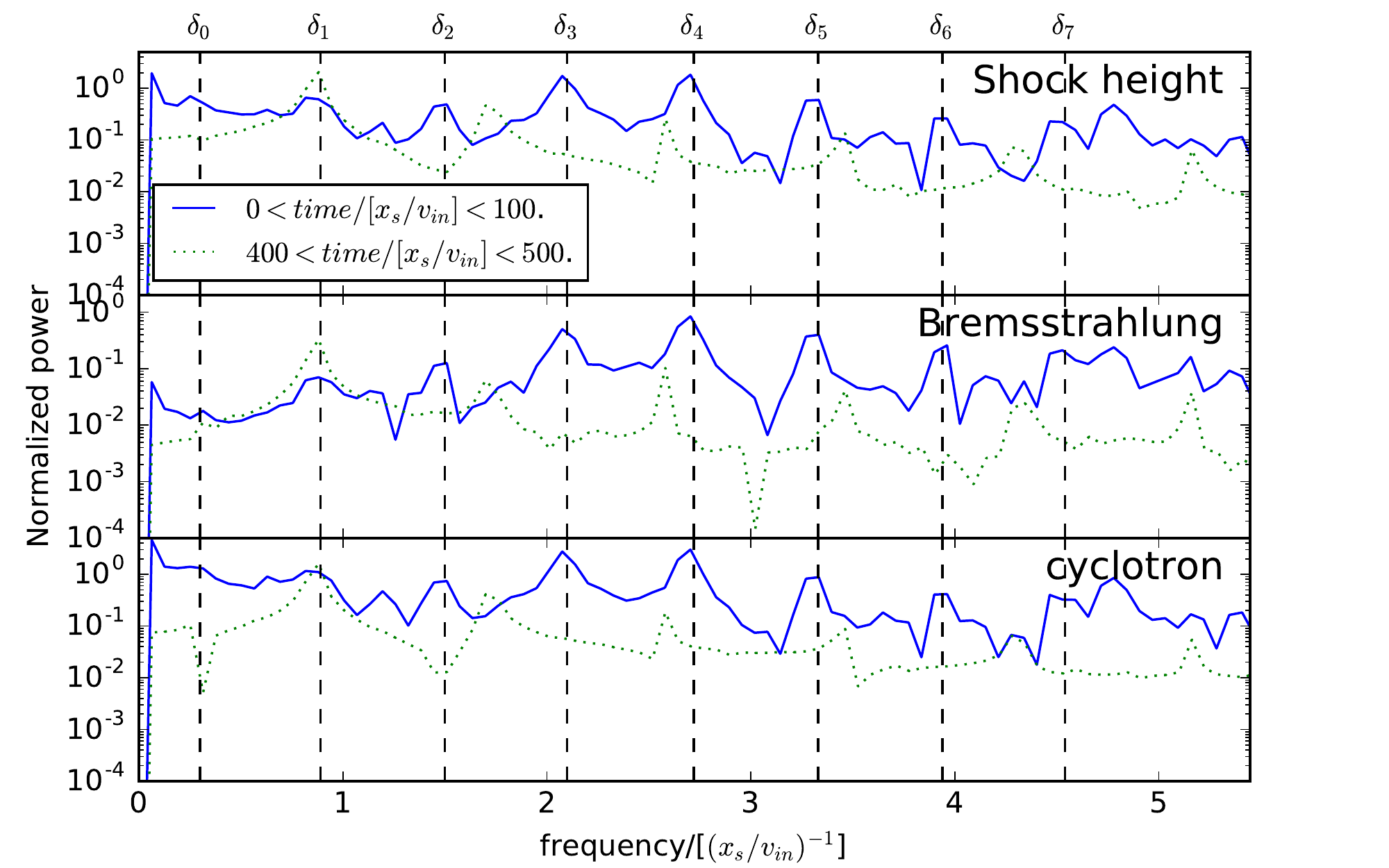}
\caption{The oscillation modes of i) shock position, ii) bremsstrahlung radiation and iii) cyclotron radiation of a bremsstrahlung dominated ($\epsilon_s \simeq 0$) accretion shock for the time range $0<t/[x_s/v_{in}]<100$ and $400<t/[x_s/v_{in}]<500$. The normal mode frequencies ($\delta_0, \delta_1,..$) obtained from the linear analysis are represented by the vertical dashed lines which are seen to match the mode frequencies in early evolution as we start from the steady solution with minimum perturbation.}
\label{oscillation_modes_100_400}
\end{figure}

The cyclic variations of physical parameters of the post shock region as seen above should give rise to similar variations of the radiative luminosity as well. We compute the total cyclotron and bremsstrahlung emission from the post-shock region at each time step to obtain the amplitude of such time variations. The characteristic modes of this variation are then calculated from the time series data. Fig.~\ref{oscillation_modes_100_400} shows the oscillation modes of the shock front and of the radiative luminosity. The modes of variation of the bremsstrahlung and cyclotron luminosity are consistent with those of the shock position. The evolution of the accretion shock started from the equilibrium steady state. At an early stage of evolution, the mode frequencies match with those obtained from linear perturbation calculations. Later in the nonlinear evolution the mode frequencies begin to deviate from their values in the linear phase.

\begin{figure}
\centering
\includegraphics[width=0.47\textwidth]{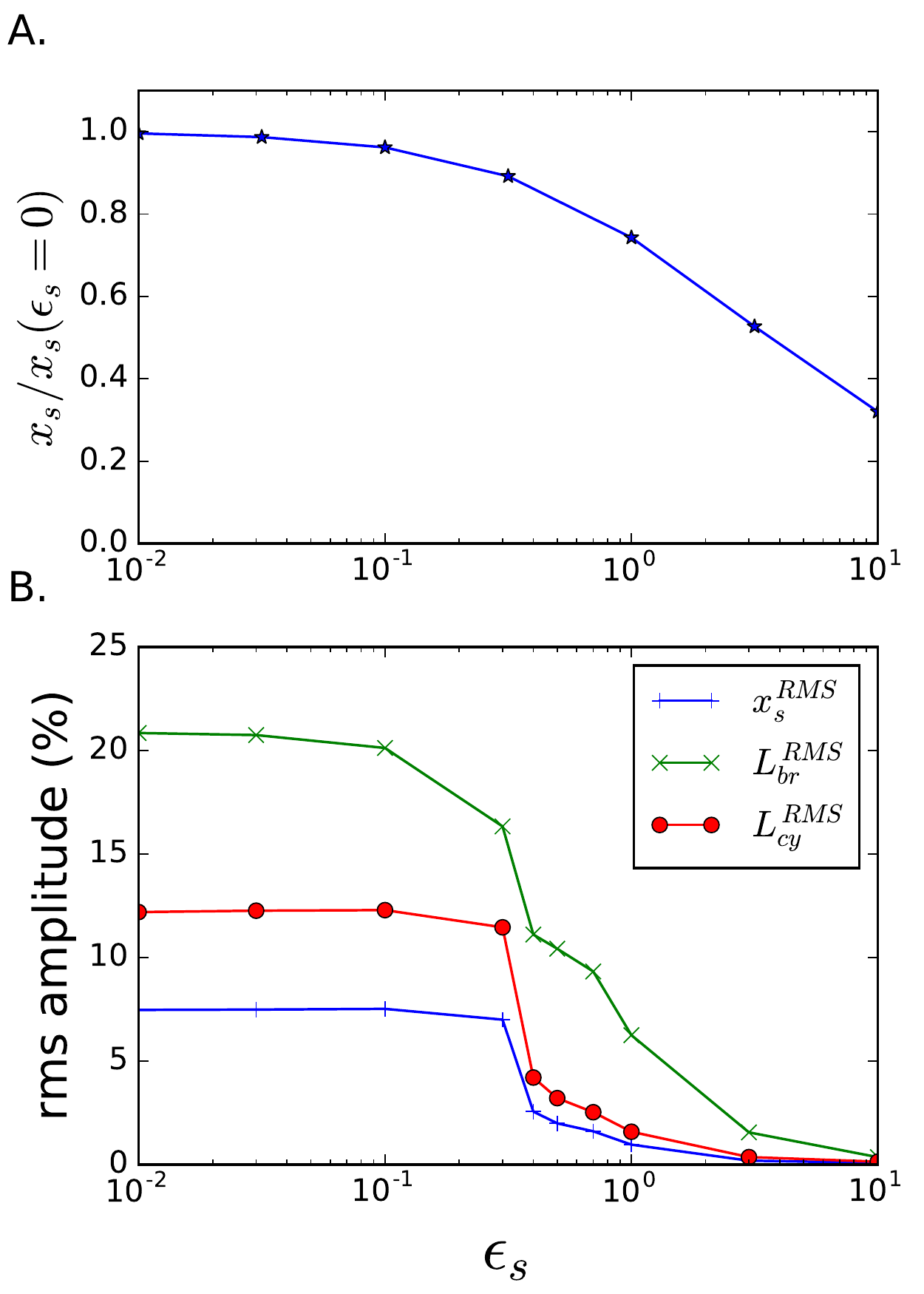}
\caption{\textbf{A.} The shock position ($x_s$) as a function of $\epsilon_s$ for a fixed $\rho_{in}$ and $v_{in}$. \textbf{B.} The dependence of the variability rms of the shock position ($x_s$), bremsstrahlung ($L_{br}$) and cyclotron ($L_{cy}$) emission on the relative strength of the cyclotron process. The rms values are calculated from the time series data of the non-linear evolution when the amplitude of the variability becomes steady.}
\label{rms_amplitude}
\end{figure}

The dependence of the dynamical behaviour of the shocked matter on the relative strength of cyclotron emission ($\epsilon_s$) is displayed in Fig.~\ref{rms_amplitude}. For a fixed inflow density ($\rho_{in}$) and velocity ($v_{in}$), cyclotron emission enhances the effective cooling, causing a reduction in the shock height (Fig.~\ref{rms_amplitude}A). Fig.~\ref{rms_amplitude}B shows the variation of the rms amplitudes of the shock position and the cooling rates with $\epsilon_s$. For a given inflow condition, the rms amplitudes drop with increasing $\epsilon_s$, tending to zero for $\epsilon_s>1$.

\begin{figure}
\centering
\includegraphics[width=0.47\textwidth]{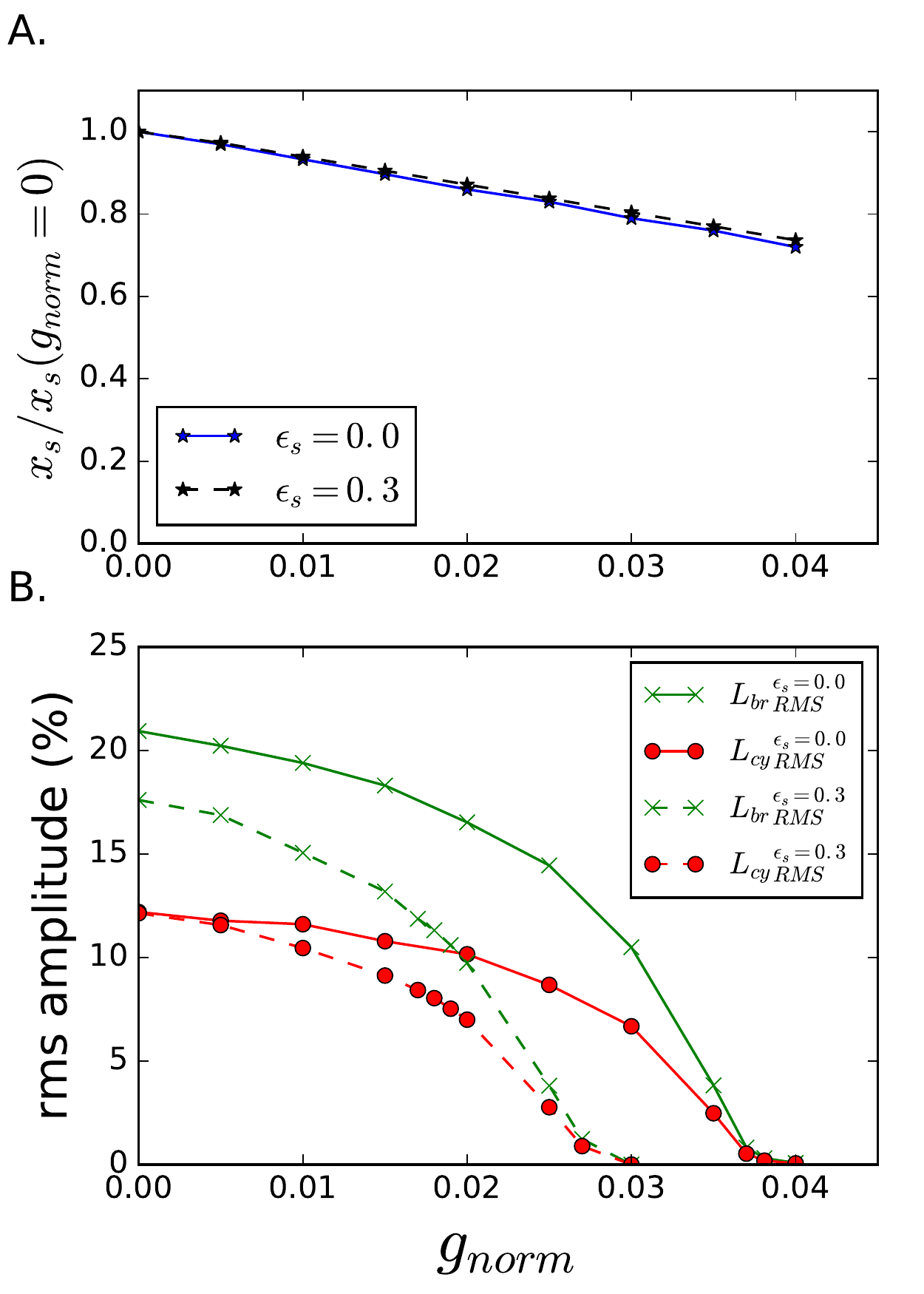}
\caption{\textbf{A.} The shock position ($x_s$) as a function of $g_{norm}$ for fixed values of $\rho_{in}$, $v_{in}$ and $\epsilon_s$. \textbf{B.} The dependence of the variability rms of the shock position, bremsstrahlung and cyclotron emission on the relative strength of gravitational acceleration.}
\label{g_rms_amplitude}
\end{figure}

Inclusion of gravitational acceleration imposes a stratification on the equilibrium structure and this influences the nonlinear dynamical evolution. From the nonlinear study of stratified accretion columns using $\textsc{pluto}$, we find that the mode frequencies remain almost unchanged but the rms amplitude of the variability changes. As  mentioned in section~\ref{subsec_eq_sol}, the gravitational stratification due to a uniform acceleration $g_{norm}$ reduces the shock height which varies almost linearly with $g_{norm}$ (Fig.~\ref{g_rms_amplitude}A). This $g_{norm}$-dependence of shock height remains valid even for different values of cyclotron cooling fraction. Fig.~\ref{g_rms_amplitude}B shows that the rms amplitude decreases for stronger gravitational acceleration and the variability vanishes beyond a threshold value of $g_{norm}$. In the presence of a strong uniform gravitational acceleration, as the plasma is compressed at the base of the column, the excess kinetic energy (converted from the gravitational potential energy) contributes additional thermal pressure. The steady solution presented in Fig.~\ref{shock_equilibrium}A shows in this case a relatively narrower region satisfying the criteria ($t_{sound}>t_{brem}$) to form a secondary shock. This reduces the strength of the secondary shock in general and, above a certain value of gravitational acceleration, prevents the formation of the secondary shock altogether. If the strength of the gravitational acceleration is not very strong, but it reduces significantly over the vertical height of the post-shock region, then the secondary shock generated at the base may gain additional strength (as a second-order effect) as it reaches the less stratified region near the primary shock.

We compare these predicted dynamical characteristics of post-shock accretion column with observations in section~\ref{observations_comparison}.

\subsection{Accumulated matter and magnetic field geometry}
\label{sec:mound}
\begin{figure}
\centering
\includegraphics[width=0.47\textwidth]{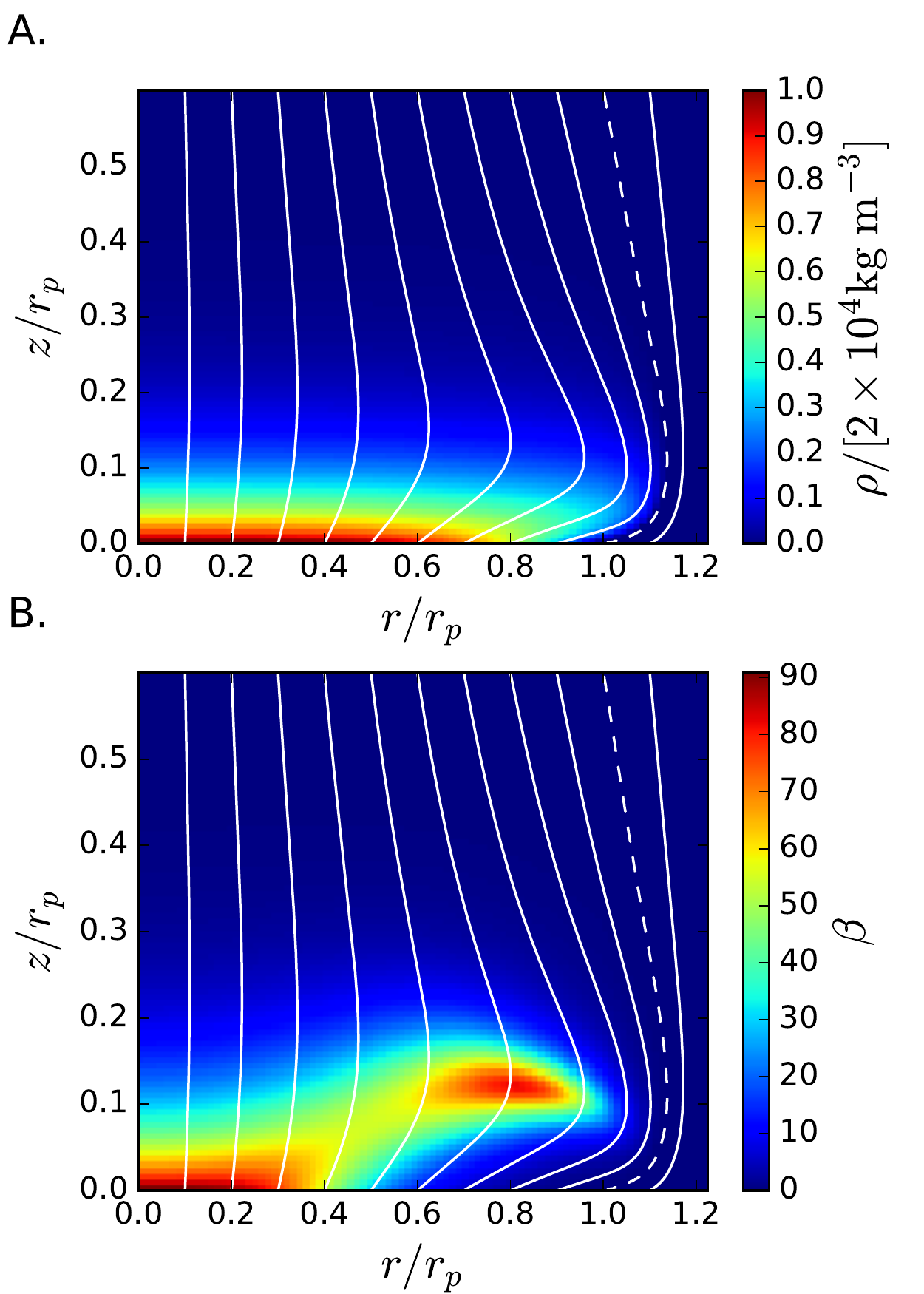}
\caption{The axisymmetric isothermal profile of \textbf{A.} the matter density ($\rho$) and \textbf{B.} the plasma beta ($\beta$) for accumulated matter at the base of an accretion column. The pressure scale height is $0.07$ times the polar cap radius. Here the dashed line (- -) indicates the outer boundary of the accumulated matter and the continous white lines represent the magnetic field lines. Distortion of the parallel field lines is observed near the outer boundary. Here we assume that the isothermal plasma at $10^6$~K is accumulated on the surface of a 0.66~M$_\odot$ white dwarf and distributed over the polar-cap region with a radius $r_p$= $10^5$~m.}
\label{accre_base_rho_beta}
\end{figure}

Relatively cool and dense accreted matter accumulates at the base of the accretion column and settles into an equilibrium configuration in the presence of local magnetic pressure. We calculate the static equilibrium solution of this axisymmetric isothermal plasma profile. For a very low matter density, the vertical profile just varies exponentially with height. As the matter density increases, the magnetic field lines at the outer boundary bend. Fig.~\ref{accre_base_rho_beta}A shows such a configuration with distorted magnetic field lines near the outer edge. Any perturbation of this equilibrium plasma structure will cause a disturbance in the magnetic field. Such a disturbance would propagate at Alfv\'en speed ($v_A=\frac{B}{\sqrt{\mu_0\rho}}$), setting the characteristic Alfv\'en time scale $t_A=\frac{r_p}{v_A}=r_p\frac{\sqrt{\mu_0\rho_{av}}}{B_{av}}$; where $\rho_{av}$ and $B_{av}^2/(2\mu_0)$ are the volume averaged matter density and magnetic energy density respectively.

From the minimum energy principle it can be shown that the bending of the magnetic field lines due to loading by adiabatic accumulated matter becomes unstable when the over pressure $\beta=p/\frac{B^2}{2\mu_0}$ crosses the limit $\beta_c\sim\alpha_a\frac{r_p}{h}$ \citep{Litwin+2001}. Here, $h$ is the density scale height, defined as $h\equiv-(d \ln\rho/dz)^{-1}|_{z=0}$. The value of $\alpha_a$ is about 10 for adiabatic plasma but for a pure isothermal plasma $\alpha_a\gg10$. The profile of plasma $\beta$, shown in Fig.~\ref{accre_base_rho_beta}B indicates a region with a very high $\beta$ value and heavily distorted magnetic field lines. According to \cite{Litwin+2001}, the adiabatic evolution of this high $\beta$ region is prone to magnetic ballooning instability if the instability condition is met. Assuming that the over pressure balances the gravitational pull, the required threshold mass to trigger the instability can be expressed as 
\begin{align}
 \Delta M = \pi r_p^3\frac{\alpha_a}{gh}\frac{B^2}{2\mu_0} = \pi r_p^3\alpha_a\rho/\beta.
\end{align}
Hence the time scale of the ballooning instability is $t_{bi}=\frac{\Delta M}{\dot{m}}$, i.e. the required time to accumulate a mass $\Delta M$ at the accretion rate $\dot{m}$.

\begin{figure}
\centering
\includegraphics[width=0.47\textwidth]{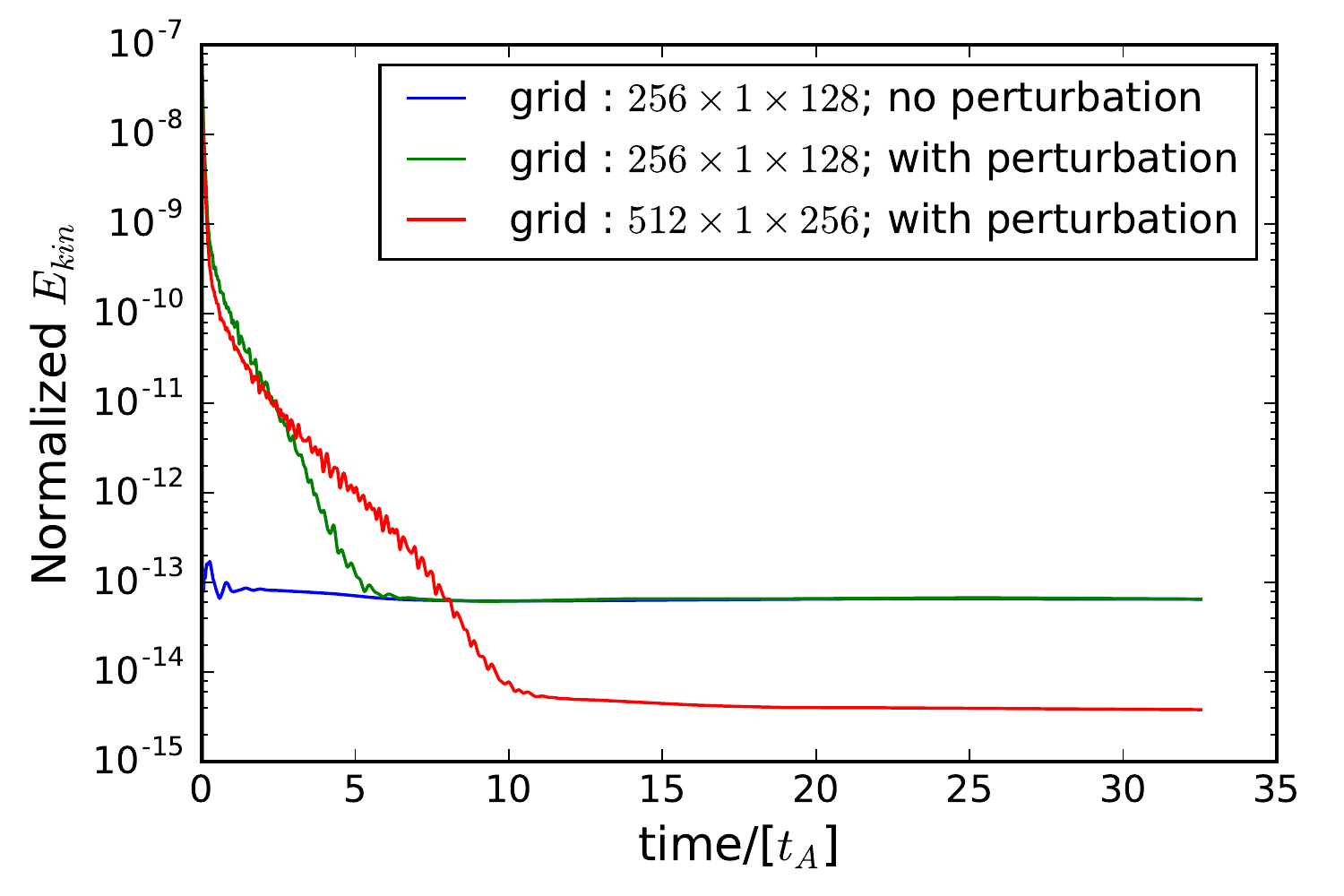}
\caption{Time evolution of the region of accumulated mass (Fig.~\ref{accre_base_rho_beta}), with and without perturbation. The unperturbed profile evolves with a small amount of kinetic energy introduced by numerical errors, while the perturbed profile settles to the equilibrium solution within a few Alfv\'en times ($t_A$). The kinetic energy added by the perturbation is converted to other forms such as the thermal and gravitational energies, and is eventually radiated away.}
\label{accre_base_instability}
\end{figure}

To study the stability characteristics of the accumulated mass, we evolve the equilibrium solution adiabatically with some perturbations using \textsc{pluto} MHD code \citep{Mukherjee+2013a, Mukherjee+2013b}. For the adiabatic evolution, we consider the adiabatic index of plasma to be $\gamma=\frac{5}{3}$. The dynamical  evolution of the distorted magnetic field region of Fig.~\ref{accre_base_rho_beta} exhibits stability for the unperturbed profile while perturbed systems settle to the equilibrium configuration in Alfv\'en time-scale (Fig.~\ref{accre_base_instability}). A small perturbation in matter density is added to the equilibrium solution which builds up kinetic energy at the initial stage. As time passes, the kinetic energy is converted to thermal, gravitational and magnetic energies. The evolution towards the equilibrium state is a direct consequence of the isothermal nature of this region, even for heavier accumulated plasma. From our low-resolution study, we do not identify any prominent oscillatory behaviour, but high-resolution study of a more detailed model may find detectable oscillatory behaviour of the perturbed configuration \citep{Payne+Melatos2007}.

\begin{figure}
\centering
\includegraphics[width=0.47\textwidth]{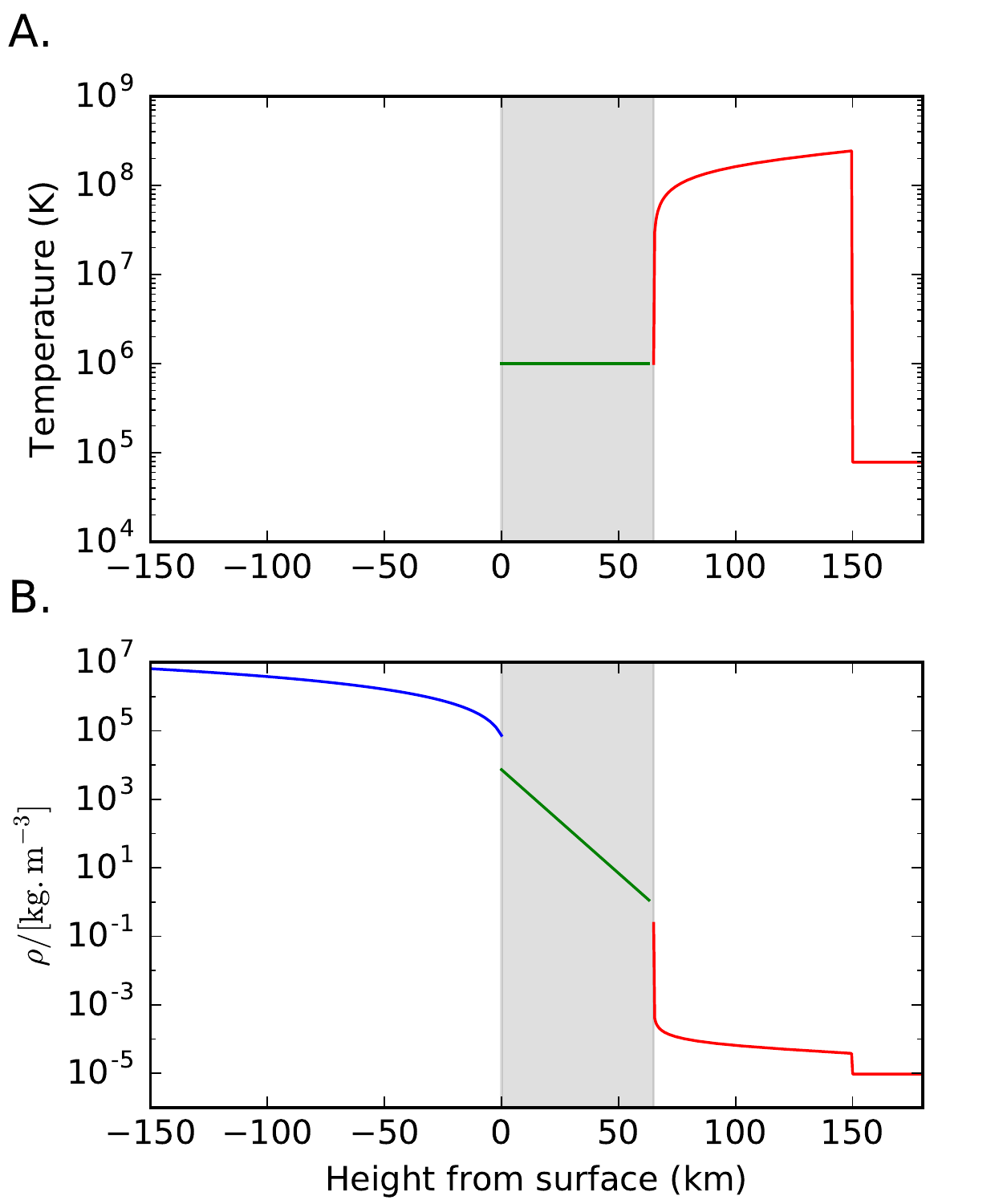}
\caption{The geometrical location of the isothermal accumulated matter (Fig.~\ref{accre_base_rho_beta}) on the white dwarf surface is shown shaded in grey. The post-shock accretion region stays above this accumulated matter. The temperature and the density profile along the axis of the polar region are shown here. The height of the accumulated mass region is assumed to smoothly match the density at the base of the accretion column.}
\label{accre_base_surface}
\end{figure}
\section{Comparison with observations and discussion}\label{observations_comparison}

In the previous sections, we have presented a model of the post-shock accretion column and its dynamical properties. Here we compare these results with the observed characteristics of the polar V834~Cen.

\subsubsection*{V834~Centauri :} The polar V834~Cen is one of the sources that exhibit QPOs with frequency about a Hz in some optical observations. From the spectroscopic ephemeris, \cite{Schwope+1993} estimate the mass of the white dwarf to be $0.66_{-0.16}^{+0.19}$~M$_\odot$. Bremsstrahlung fits to the observed X-ray spectra of this source yield a white dwarf mass estimate of 0.5~M$_\odot$ with a larger range of uncertainty \citep{Cropper+1998}. Observed cyclotron and Zeeman features indicate a stellar magnetic field of strength $2.3\times10^3$~T \citep{Schwope+Beuermann1990}. From optical observations carried out in 1977, \cite{Middleditch1982} identified a quasi-periodic oscillation with an rms amplitude of 1.2\% in the frequency range $0.4-0.8$~Hz. Recently, from VLT-ULTRACAM observations, \cite{Mouchet+2017} also find QPOs in all three optical bands ($r', \rm{He II ~\lambda4686}, u'$) with frequency $0.4-0.8$~Hz and rms amplitudes of 2.1\%, 1.5\% and 0.6\% respectively. Here the $r'$ band power spectrum shows QPO over a broader frequency range ($0.25-5$~Hz). Despite the identification of QPOs in different optical observations, there is no specific detection of the QPO in the X-ray energy band. \cite{Imamura+2000} placed an upper bound of 14\% on the X-ray QPO amplitude while simultaneous optical observations showed  QPOs at $\sim$1~Hz with modulation amplitude of $\sim$1\%. \cite{Bonnet-Bidaud+2015} estimate the upper bound on the X-ray rms variation to be less than $9\%$ based on XMM-Newton archival data. \cite{Imamura+1991} studied the dynamics of the post shock accretion column on a 0.3 M$_\odot$ white dwarf and attributed the 1-3~s variability in the optical emission to a limit cycle oscillation of the column.
 
 \begin{table*}
 \caption{The accretion column structure and the expected variability of V834 Cen with a  white dwarf mass 0.66~M$_\odot$, stellar radius $8.315\times10^6$~m and polar magnetic field (B) $2.3\times10^3$~T for different polar cap radii. The accretion rate is considered to be $1.4\times10^{12}$~kg/s to generate a luminosity $1.5\times10^{25}~\rm{J~s^{-1}}$. The lower bound of the range of the QPO frequency corresponds to the fundamental of the linear modes.}
 \begin{supertabular}{c|c c c c c c c}
  \hline
	 \bf Variables	& $r_p=3\times10^4$ m	& $r_p=10^5$ m		& $r_p=1.4\times10^5$ m		& $r_p=2\times10^5$ m	& $r_p=3\times10^5$ m	& $r_p=5\times10^5$ m	\\
  \hline
  	 Area (m$^2$)	&$2.83\times10^{9}$	&$3.14\times10^{10}$	&$6.16\times10^{10}$	& $1.26\times10^{11}$	& $2.83\times10^{11}$	& $7.85\times10^{11}$	\\
	 $\rho_{in}~(\rm kg.m^{-3})$& $1.06\times10^{-4}$& $9.65\times10^{-6}$& $4.95\times10^{-6}$&$2.39\times10^{-6}$&$1.06\times10^{-6}$&$3.99\times10^{-7}$\\
 \hline
%	\multirow{3}{*}{no gravity}
	$x_s$ (m)	& $9.049\times10^4$	& $1.262\times10^5$	& $2.165\times10^5$	& $3.51\times10^5$	& $5.69\times10^5$	& $6.93\times10^5$	\\
	$x_s/R_{wd}(\%)$&	1.112		&	1.551		&	2.66		& 4.55			& 7.0			& 8.52	\\
	$x_s/v_{in}$(s)&$2.77\times10^{-3}$	& $2.7\times10^{-2}$	& $4.7\times10^{-2}$	& 0.0814		& 0.127			& 0.155\\
  \hline
	 $\epsilon_s$	&	$10^{-2}$	&	0.148		&	0.373		& 0.978			& 2.89			& 11.94\\
	 $g_{norm}$	& $8.\times10^{-4}$& $7.75\times10^{-3}$	& $1.33\times10^{-2}$	& $2.57\times10^{-2}$	& $4.04\times10^{-2}$	& $4.26\times10^{-2}$\\
%	$b_{norm}$	&	43.7		&	145.7		&	204		&	291.4		&	437.14		\\
  \hline
	 $L_{cycl}~(\rm J/s)$& $1.25\times10^{22}$& $3.32\times10^{23}$	& $7.4\times10^{23}$	& $1.62\times10^{24}$	& $3.19\times10^{24}$	& $4.98\times10^{24}$\\
	 $L_{brem}~(\rm J/s)$& $1.34\times10^{25}$& $1.23\times10^{25}$	& $1.12\times10^{25}$	& $1.02\times10^{25}$	& $8.08\times10^{24}$	& $4.60\times10^{24}$\\
	 $L_{cy}^{RMS}~(\%)$&		12.2	&	11.5		&	10.25		&	0.07		&	$\sim$0.0	& $\sim$0.0\\
	 $L_{br}^{RMS}~(\%)$&		20.6	&	18.1		&	14.4		&	0.12		&	$\sim$0.0	& $\sim$0.0\\
	 QPO freq. (Hz)	&	18-50		&	1.8-5.0		&	1.0-2.9		&	0.6-1.7		&	0.4-1.1	& 0.3-0.9\\
  \hline
 \end{supertabular}\label{table_V834_qpo_rms}
\end{table*}

To model the accretion column, we consider the magnetic white dwarf to have a mass 0.66~M$_\odot$ and a polar magnetic field strength $2.3\times10^3$~T. The mass accretion rate is assumed to be $1.4\times10^{12}~\rm{kg~s^{-1}}$, corresponding to the observed X-ray luminosity $1.5\times10^{25}~\rm{J~s^{-1}}$ \citep{Bonnet-Bidaud+2015}. This inflow rate is over the entire polar cap region and there is no direct observational handle on the polar cap surface area. However since our model of the accretion column requires an inflow rate per unit area, we need an estimate of the polar cap radius $r_p$ to proceed. An estimate using the Alfv\'en radius for a dipolar field of strength $2.3\times10^3$~T, gives $r_p\sim5\times10^5~\rm{m}$. We calculate the steady post-shock configurations for this source assuming different values of $r_p$ ranging from $3\times10^4~\rm{m}$ to $5\times10^5~\rm{m}$.  The specific accretion rate (and hence $\rho_{in}$) decreases as the polar cap radius is increased for a fixed total accretion rate as shown in table~\ref{table_V834_qpo_rms}. As the inflow matter density ($\rho_{in}$) decreases the shock height increases from $9\times10^4~\rm{m}$ to $5.7\times10^5~\rm{m}$, growing to about 1-7\% of the white dwarf radius. The effects of gravitational acceleration ($g_{norm}$) and the variation of the gravitational force along the accretion flow become important for these taller accretion columns. We also find that the ratio between the cyclotron emission rate to the bremsstrahlung rate (i.e. $\epsilon_s$) increases as the specific accretion rate decreases.

We next study the dynamical evolution of these steady solutions using \textsc{pluto}, as in sec.~\ref{subsec:perturb_nonlin}. Starting from the steady solution, as time passes, perturbations appear in the post-shock region. The shock position as well as the emission rates due to cyclotron and bremsstrahlung processes show an oscillatory behaviour. The rms and the frequency range of the oscillations for different shock structures are listed in Table~\ref{table_V834_qpo_rms}. If the polar cap radius ($r_p$) is greater than $1\times10^5~\rm{m}$, then the rms becomes very small as such columns are strongly affected by cyclotron cooling and gravitational stratification. QPO frequencies in the observed range ($0.4-0.8$~Hz) can be obtained for an accretion column with polar cap radius $r_p\sim 2\times10^5~\rm{m}$. The corresponding oscillation rms is predicted to be very low ($<1\%$) in both cyclotron and bremsstrahlung luminosity. The observed optical rms of about 1-2\% requires a polar cap radius $r_p=1.6\times10^5~\rm{m}$ for which the QPO frequencies are a little higher, in the range $0.9-2.4$~Hz. The corresponding variations of bremsstrahlung X-ray luminosity remain weaker than the observed upper bound of 9\% rms. So we find that the observed QPO characteristics are nearly consistent with the theoretical predictions for such a post-shock column.
 
 \begin{figure}
\centering
\includegraphics[width=0.47\textwidth]{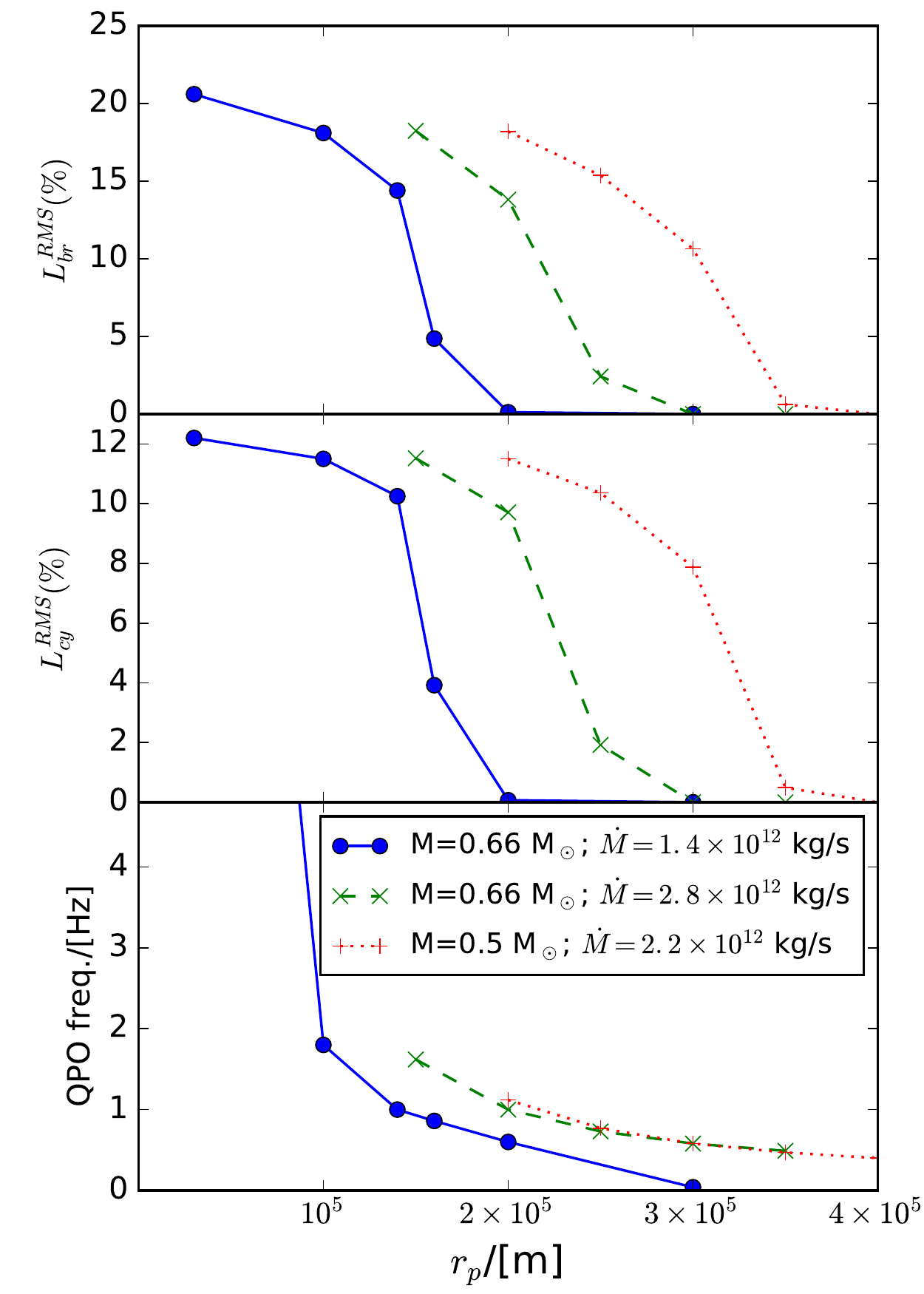}
\caption{The surface area dependency of the variability rms of bremsstrahlung, cyclotron emission and the fundamental QPO frequency ($\delta_0$) for different post-shock parameter values. The polar magnetic field strength of the white dwarf is taken to be $2.3\times10^3$~T for all the cases.}
\label{V834Cen_1d}
\end{figure}

The real parameters of the V834~Cen system, may, however be different from the nominal values chosen above. The mass estimate of the white dwarf has a significant uncertainty, and so does the distance estimate. Taken together, they admit a rather large range of the mass accretion rate. A higher accretion rate would end up reducing the shock height and hence the dominance of cyclotron cooling. A lower mass of the white dwarf would, in turn, reduce the gravitational stratification and strengthen the cooling catastrophe.  For a fixed polar cap area, both these will go towards increasing the expected QPO rms and the QPO frequency by different amounts (see Fig.~\ref{V834Cen_1d}). A simultaneous measurement of the QPO frequency and rms at optical and X-ray bands can therefore be used to place important constraints on the system parameters.  

As discussed in sec.~\ref{sec:mound}, the accumulated matter at the base of the accretion column may potentially trigger a magnetic instability, although our computations did not reveal any unstable behaviour. In case of the column model of V834~Cen, we have extended these computations to a total accumulated mass amounting to $10^{18}$~kg, but yet  do not find any instability. Hence, the ballooning instability time scale appears to be much longer ($>10^6$~s) than the QPO period. However in this case the Alfv\'en time is in the range $\sim 1-10$~s, so resonant interaction of magnetic modes with thermally driven oscillations of the overlying column may enhance the QPO amplitude. Such a study requires a much more elaborate computational set up, and is beyond the scope of the present work.

We therefore find that thermal instability in the radiating post-shock accretion column can nearly reproduce the characteristics of the Hz-frequency QPOs observed from the polars. This conclusion is at variance with that of \cite{Mouchet+2017}, who predict a much higher QPO frequency ($\sim19~\rm{Hz}$) from their model, for very similar input parameters (stellar mass, magnetic field and accretion rate). The main differences between their work and ours are in: i) the expression of $\epsilon_s$ (equation~\ref{expression_epsilon_s}) ii) the lower boundary condition and iii) the initial setup. The value of $\epsilon_s$ decides the relative importance of  cyclotron cooling and hence the height of the post-shock region (Fig.~\ref{rms_amplitude}A), which in turn determines the characteristic time scale. The expression of $\epsilon_s$ that we have used here is consistent with that of \cite{Wu+1994} and its derivation is explained above (eq.~\ref{eq_brem_nu}--\ref{expression_epsilon_s}). The lower boundary condition hardly influences the oscillation frequency but has an important effect on the strength of the variability. We use an absorbing boundary condition at the base of the column which minimises reflection and hence has a stabilising effect. However this can be discerned only at very low variability amplitudes \citep{Mignone2005}. Although there are some differences in the initial setup used by us compared to that of \cite{Mouchet+2017}, the general characteristics of the variability at late times is expected to be independent of the initial setup. The main contributor to the difference between our results and those of \cite{Mouchet+2017} therefore appears to be the expression of $\epsilon_s$.

To model the post-shock accretion column, we have considered here only the one-dimensional vertical variations along the magnetic field. In reality, there may be spatial variation in the mass flow rate across the polar cap. Although strong vertical magnetic field restricts the horizontal motion of the plasma, dynamics of the column would be affected by the spatial variation in column structure and  transverse modes of disturbance may possibly be excited \citep{Strickland+Blondin1995}. These may be capable of influencing the post-shock dynamics and hence the variability in the observables.

\section{Conclusions} \label{section_conclusion}

Using an one-dimensional model of the post-shock accretion column on a polar, we estimated the range of  variability time scales and amplitudes via numerical computation of its dynamical evolution. The predicted variability is consistent with that observed in the polar V834~Cen. Our primary results are summarised below.
\begin{enumerate}
 \item In a radiative post-shock accretion column, localised efficient cooling forms a weak ($\mathcal{M}\lesssim 1$) secondary shock. In the case of pure bremsstrahlung emission, the cooling is more active at the base of the column, and hence a secondary shock is formed there. This secondary shock propagates towards the primary shock from the base and reduces the shock height until the pressure of matter thus compressed starts to drive an expansion. When the height of the primary shock increases from the compressed state, the reduced pressure of the high density plasma at the base generates another secondary shock to continue a new cycle. This quasi-periodic dynamical oscillation of the post-shock region manifests in the variability of the emitted radiation.
 \item The presence of cyclotron emission, which dominates in the high temperature region near the primary shock front, reduces the amplitude of the variability as the effect of cooling becomes significant over the entire post-shock region. 
 \item Under the influence of the gravity of the white dwarf, vertical stratification reduces the height of post-shock region. The formation of the dense region at the base is restricted as a higher thermal pressure is created at the expense of gravitational potential. This weakens the secondary shock and consequently reduces the variability amplitude.  
 \item The base of the accretion column plays an important role in driving the variability as the secondary shock is generated there. The magnetically confined, cooler accumulated matter at the base of the accretion column may distort the magnetic field lines due to the internal pressure. Perturbations of this accumulated mass may generate magnetic disturbances that could enhance the oscillations of the post-shock region.
\end{enumerate}

\section{Acknowledgement}
PB thanks CSIR, India for Research Fellowship grant SPM-09/545(0221)/2015-EMR-I. We thank David Buckley, K.P. Singh, Indranil Chattopadhyay, Emeric Falize and Andrea Mignone for useful discussions. Most of the numerical computations reported here were carried out using the IUCAA High Perfomance Computing facility.

\appendix
\section{Magnetic field structure at the region of the accreted matter accumulation}\label{GS_solution_methos}

We consider that the accumulated matter forms an axisymmetric structure at the polar cap region ($<r_p$) with uniform field strength $B_0$.
In equilibrium, the force balance equation is given as
\begin{align}
 \nabla p + \rho\nabla\Phi_g + \mathbf{j}\times\mathbf{B} = 0. \label{eq_force_balance}
\end{align}
To find the axisymmetric accumulated mass configuration in cylindrical polar coordinate $(R,~ \phi,~ z)$, co-aligned with the accretion axis, we assume that the magnetic field is axisymmetric and untwisted which can be expressed as
\begin{align}
 \mathbf{B} = \frac{\nabla\times\psi}{R}.%\\
 %\rm{Therefore,}~~ B_R = \frac{1}{R}\frac{\partial\psi}{\partial z};~~ B_z = \frac{1}{R}\frac{\partial\psi}{\partial R}.
\end{align}
The flux function $\psi=\int_0^R BRdR$ and for the unperturbed field $\psi=\frac{1}{2}B_0R^2$ with maximum at the polar cap radius $\psi_p=\frac{1}{2}B_0r_p^2$. Hence, the equation~\ref{eq_force_balance} becomes 
\begin{align}
 \nabla p + \rho\nabla\Phi_g + \frac{\Delta^2\psi}{\mu_0R^2}\nabla\psi = 0,
\end{align}
 where, $\Delta^2=R\frac{\partial}{\partial R}\left(\frac{1}{R}\frac{\partial}{\partial R}\right)+\frac{\partial^2}{\partial z^2}$. The vector components of this equation are
\begin{align}
 p_R + \frac{\Delta^2\phi}{\mu_0R^2}\psi_R = 0;\label{eq_force_comp_R}\\
 p_z - \rho g+ \frac{\Delta^2\psi}{\mu_0R^2}\psi_z = 0~~\textrm{or,}~p_z - \rho g+ \frac{-p_R}{\psi_R}\psi_z = 0. \label{eq_force_comp_z}
\end{align}
Here, $\nabla\Phi_g = - g\mathbf{\hat{z}} = |g|\mathbf{\hat{z}}$ and the subscripts $R$ and $z$ indicate the differentials with respect to these variables ($R$ and $z$ respectively).
The characteristic solution of the equation~\ref{eq_force_comp_z} can be expressed as \citep{Mukherjee+2012}
\begin{align}
 dz = \frac{dr}{-\psi_z/\psi_R} = \frac{dp}{-\rho |g|}. \label{eq_char_eqsn}
\end{align}
This characteristic equation~\ref{eq_char_eqsn} can be solved for a given equation of state (EoS). The matter density ($\rho$) distribution can be expressed as a function of the flux function $\psi(R, z)$ \citep{Payne+Melatos2004, Mukherjee+2012},
\begin{align}
   \rho(\psi)=\begin{cases}
    \left[\frac{|g|(\gamma-1)z_0(\psi)}{\gamma k_{ad}}\left(1-\frac{z}{z_0(\psi)}\right)\right]^\frac{1}{\gamma-1}, & \text{EoS $p=k_{ad}\rho^\gamma$};\\
    \rho_0(\psi)\exp\left(-\frac{|g|z}{c_s^2}\right), & \text{EoS $p=c_s^2\rho$},
  \end{cases}
\end{align}
where, $k_{ad}$ is the adiabatic constant and $c_s$ is the isothermal sound speed.
Now, the required step is to find the flux function $\psi(R,z)$ from equation~\ref{eq_force_comp_R}
\begin{align}
 \Delta^2\psi = -\mu_0R^2\frac{dp}{d\psi} = -\mu_0R^2\left(\frac{\partial p}{\partial \rho}\right)\frac{d\rho}{d\psi}. \label{eq_GS}
\end{align}
This Grad-Shafranov like equation~\ref{eq_GS} can be solved for a given functional form of $z_0(\psi)$ for adiabatic EoS (or $\rho_0(\psi)$ for isothermal EoS). Here we assume the following functional forms
\begin{align}
 z_0(\psi)/z_c=1-\left(\frac{\psi}{\psi_p}\right)^2; ~~\rm{adiabatic~ EoS},\\
 \rho_0(\psi)/\rho_c=1-\left(\frac{\psi}{\psi_p}\right)^2; ~~\rm{isothermal~ EoS}.
\end{align}
Hence, for adiabatic EoS, $z_c$ represents the maximum height of the accumulated matter whereas for isothermal EoS $\rho_c$ represents the maximum matter density at the center of the base.

We solve equation~\ref{eq_GS} by the method of successive over relaxation (SOR) as described in \cite{Press+2007}. For the boundary condition on the magnetic field we assume that $\psi$ maintains its unperturbed value at the base and also at the top level. For the dynamical evolution we use \textsc{pluto} MHD code where we evolve with perturbation a part of the cylindrical region where the magnetic field is strongly distorted. A fixed gradient boundary condition \citep{Mukherjee+2013a} is imposed at the outer radial boundary while the other boundary is kept fixed at the equilibrium values. 

%==========================================
%----------------- Bibliography and bibfile

\def\aj{AJ}%
\def\actaa{Acta Astron.}%
\def\araa{ARA\&A}%
\def\apj{ApJ}%
\def\apjl{ApJ}%
\def\apjs{ApJS}%
\def\ao{Appl.~Opt.}%
\def\apss{Ap\&SS}%
\def\aap{A\&A}% 
\def\aapr{A\&A~Rev.}%
\def\aaps{A\&AS}%
\def\azh{AZh}%
\def\baas{BAAS}%
\def\bac{Bull. astr. Inst. Czechosl.}%
\def\caa{Chinese Astron. Astrophys.}%
\def\cjaa{Chinese J. Astron. Astrophys.}%
\def\icarus{Icarus}%
\def\jcap{J. Cosmology Astropart. Phys.}%
\def\jrasc{JRASC}%
\def\mnras{MNRAS}%
\def\memras{MmRAS}%
\def\na{New A}%
\def\nar{New A Rev.}%
\def\pasa{PASA}%
\def\pra{Phys.~Rev.~A}%
\def\prb{Phys.~Rev.~B}%
\def\prc{Phys.~Rev.~C}%
\def\prd{Phys.~Rev.~D}%
\def\pre{Phys.~Rev.~E}%
\def\prl{Phys.~Rev.~Lett.}%
\def\pasp{PASP}%
\def\pasj{PASJ}%
\def\qjras{QJRAS}%2215.bib
\def\rmxaa{Rev. Mexicana Astron. Astrofis.}%
\def\skytel{S\&T}%
\def\solphys{Sol.~Phys.}%
\def\sovast{Soviet~Ast.}%
\def\siamr{SIAMR}%
\def\ssr{Space~Sci.~Rev.}%
\def\zap{ZAp}%
\def\nat{Nature}%
\def\iaucirc{IAU~Circ.}%
\def\aplett{Astrophys.~Lett.}%
\def\apspr{Astrophys.~Space~Phys.~Res.}%
\def\bain{Bull.~Astron.~Inst.~Netherlands}%
\def\fcp{Fund.~Cosmic~Phys.}%
\def\gca{Geochim.~Cosmochim.~Acta}%
\def\grl{Geophys.~Res.~Lett.}%
\def\jcp{J.~Chem.~Phys.}%
\def\jgr{J.~Geophys.~Res.}%
\def\jqsrt{J.~Quant.~Spec.~Radiat.~Transf.}%
\def\memsai{Mem.~Soc.~Astron.~Italiana}%
\def\nphysa{Nucl.~Phys.~A}%
\def\physrep{Phys.~Rep.}%
\def\physscr{Phys.~Scr}%
\def\planss{Planet.~Space~Sci.}%
\def\procspie{Proc.~SPIED}%
\let\astap=\aap
\let\apjlett=\apjl
\let\apjsupp=\apjs
\let\applopt=\ao

\bibliographystyle{mnras}	% (uses file "plain.bst")

\bibliography{ref.bib}

\begin{thebibliography}{}
\makeatletter
\relax
\def\mn@urlcharsother{\let\do\@makeother \do\$\do\&\do\#\do\^\do\_\do\%\do\~}
\def\mn@doi{\begingroup\mn@urlcharsother \@ifnextchar [ {\mn@doi@}
  {\mn@doi@[]}}
\def\mn@doi@[#1]#2{\def\@tempa{#1}\ifx\@tempa\@empty \href
  {http://dx.doi.org/#2} {doi:#2}\else \href {http://dx.doi.org/#2} {#1}\fi
  \endgroup}
\def\mn@eprint#1#2{\mn@eprint@#1:#2::\@nil}
\def\mn@eprint@arXiv#1{\href {http://arxiv.org/abs/#1} {{\tt arXiv:#1}}}
\def\mn@eprint@dblp#1{\href {http://dblp.uni-trier.de/rec/bibtex/#1.xml}
  {dblp:#1}}
\def\mn@eprint@#1:#2:#3:#4\@nil{\def\@tempa {#1}\def\@tempb {#2}\def\@tempc
  {#3}\ifx \@tempc \@empty \let \@tempc \@tempb \let \@tempb \@tempa \fi \ifx
  \@tempb \@empty \def\@tempb {arXiv}\fi \@ifundefined
  {mn@eprint@\@tempb}{\@tempb:\@tempc}{\expandafter \expandafter \csname
  mn@eprint@\@tempb\endcsname \expandafter{\@tempc}}}

\bibitem[\protect\citeauthoryear{{Aizu}}{{Aizu}}{1973}]{Aizu1973}
{Aizu} K.,  1973, \mn@doi [Progress of Theoretical Physics]
  {10.1143/PTP.49.1184}, \href
  {http://adsabs.harvard.edu/abs/1973PThPh..49.1184A} {49, 1184}

\bibitem[\protect\citeauthoryear{{Barrett} \& {Chanmugam}}{{Barrett} \&
  {Chanmugam}}{1985}]{Barrett+Chanmugam1985}
{Barrett} P.~E.,  {Chanmugam} G.,  1985, \mn@doi [\apj] {10.1086/163655}, \href
  {http://adsabs.harvard.edu/abs/1985ApJ...298..743B} {298, 743}

\bibitem[\protect\citeauthoryear{{Beardmore} \& {Osborne}}{{Beardmore} \&
  {Osborne}}{1997}]{Beardmore+Osborne1997a}
{Beardmore} A.~P.,  {Osborne} J.~P.,  1997, \mn@doi [\mnras]
  {10.1093/mnras/286.1.77}, \href
  {http://adsabs.harvard.edu/abs/1997MNRAS.286...77B} {286, 77}

\bibitem[\protect\citeauthoryear{{Bonnet-Bidaud}, {Mouchet}, {Somova}  \&
  {Somov}}{{Bonnet-Bidaud} et~al.}{1996}]{Bonnet-Bidaud+1996}
{Bonnet-Bidaud} J.~M.,  {Mouchet} M.,  {Somova} T.~A.,   {Somov} N.~N.,  1996,
  \aap, \href {http://adsabs.harvard.edu/abs/1996A%26A...306..199B} {306, 199}

\bibitem[\protect\citeauthoryear{{Bonnet-Bidaud}, {Mouchet}, {Busschaert},
  {Falize}  \& {Michaut}}{{Bonnet-Bidaud} et~al.}{2015}]{Bonnet-Bidaud+2015}
{Bonnet-Bidaud} J.~M.,  {Mouchet} M.,  {Busschaert} C.,  {Falize} E.,
  {Michaut} C.,  2015, \mn@doi [\aap] {10.1051/0004-6361/201425482}, \href
  {http://adsabs.harvard.edu/abs/2015A%26A...579A..24B} {579, A24}

\bibitem[\protect\citeauthoryear{{Brown} \& {Bildsten}}{{Brown} \&
  {Bildsten}}{1998}]{Brown+Bildsten1998}
{Brown} E.~F.,  {Bildsten} L.,  1998, \mn@doi [\apj] {10.1086/305419}, \href
  {http://adsabs.harvard.edu/abs/1998ApJ...496..915B} {496, 915}

\bibitem[\protect\citeauthoryear{{Busschaert}, {Falize}, {Michaut},
  {Bonnet-Bidaud}  \& {Mouchet}}{{Busschaert} et~al.}{2015}]{Busschaert+2015}
{Busschaert} C.,  {Falize} {\'E}.,  {Michaut} C.,  {Bonnet-Bidaud} J.-M.,
  {Mouchet} M.,  2015, \mn@doi [\aap] {10.1051/0004-6361/201425483}, \href
  {http://adsabs.harvard.edu/abs/2015A%26A...579A..25B} {579, A25}

\bibitem[\protect\citeauthoryear{{Chanmugam}}{{Chanmugam}}{1995}]{Chanmugam1995}
{Chanmugam} G.,  1995, in {Buckley} D.~A.~H.,  {Warner} B.,  eds,  Astronomical
  Society of the Pacific Conference Series Vol. 85, Magnetic Cataclysmic
  Variables. p.~317

\bibitem[\protect\citeauthoryear{{Chanmugam} \& {Wagner}}{{Chanmugam} \&
  {Wagner}}{1979}]{Chanmugam+Wagner1979}
{Chanmugam} G.,  {Wagner} R.~L.,  1979, \mn@doi [\apj] {10.1086/157352}, \href
  {http://adsabs.harvard.edu/abs/1979ApJ...232..895C} {232, 895}

\bibitem[\protect\citeauthoryear{{Chanmugam}, {Langer}  \&
  {Shaviv}}{{Chanmugam} et~al.}{1985}]{Chanmugam+1985}
{Chanmugam} G.,  {Langer} S.~H.,   {Shaviv} G.,  1985, \mn@doi [\apjl]
  {10.1086/184586}, \href {http://adsabs.harvard.edu/abs/1985ApJ...299L..87C}
  {299, L87}

\bibitem[\protect\citeauthoryear{{Chevalier} \& {Imamura}}{{Chevalier} \&
  {Imamura}}{1982}]{Chevalier+Imamura1982}
{Chevalier} R.~A.,  {Imamura} J.~N.,  1982, \mn@doi [\apj] {10.1086/160364},
  \href {http://adsabs.harvard.edu/abs/1982ApJ...261..543C} {261, 543}

\bibitem[\protect\citeauthoryear{{Cropper}, {Ramsay}  \& {Wu}}{{Cropper}
  et~al.}{1998}]{Cropper+1998}
{Cropper} M.,  {Ramsay} G.,   {Wu} K.,  1998, \mn@doi [\mnras]
  {10.1046/j.1365-8711.1998.00610.x}, \href
  {http://ukads.nottingham.ac.uk/abs/1998MNRAS.293..222C} {293, 222}

\bibitem[\protect\citeauthoryear{{Cropper}, {Wu}, {Ramsay}  \&
  {Kocabiyik}}{{Cropper} et~al.}{1999}]{Cropper+1999}
{Cropper} M.,  {Wu} K.,  {Ramsay} G.,   {Kocabiyik} A.,  1999, \mn@doi [\mnras]
  {10.1046/j.1365-8711.1999.02570.x}, \href
  {http://adsabs.harvard.edu/abs/1999MNRAS.306..684C} {306, 684}

\bibitem[\protect\citeauthoryear{{Cross} et~al.,}{{Cross}
  et~al.}{2016}]{Cross+2016}
{Cross} J.~E.,  et~al., 2016, \mn@doi [Nature Communications]
  {10.1038/ncomms11899}, \href
  {http://adsabs.harvard.edu/abs/2016NatCo...711899C} {7, 11899}

\bibitem[\protect\citeauthoryear{{Dougal} \& {Goldstein}}{{Dougal} \&
  {Goldstein}}{1958}]{Dougal+Goldstein1958}
{Dougal} A.~A.,  {Goldstein} L.,  1958, \mn@doi [Physical Review]
  {10.1103/PhysRev.109.615}, \href
  {http://adsabs.harvard.edu/abs/1958PhRv..109..615D} {109, 615}

\bibitem[\protect\citeauthoryear{{Fabian}, {Pringle}  \& {Rees}}{{Fabian}
  et~al.}{1976}]{Fabian+1976}
{Fabian} A.~C.,  {Pringle} J.~E.,   {Rees} M.~J.,  1976, \mn@doi [\mnras]
  {10.1093/mnras/175.1.43}, \href
  {http://adsabs.harvard.edu/abs/1976MNRAS.175...43F} {175, 43}

\bibitem[\protect\citeauthoryear{{Falize} et~al.,}{{Falize}
  et~al.}{2011}]{Falize+2011}
{Falize} {\'E}.,  et~al., 2011, \mn@doi [\apss] {10.1007/s10509-011-0655-4},
  \href {http://adsabs.harvard.edu/abs/2011Ap%26SS.336...81F} {336, 81}

\bibitem[\protect\citeauthoryear{{Falle}}{{Falle}}{1981}]{Falle1981}
{Falle} S.~A.~E.~G.,  1981, \mn@doi [\mnras] {10.1093/mnras/195.4.1011}, \href
  {http://adsabs.harvard.edu/abs/1981MNRAS.195.1011F} {195, 1011}

\bibitem[\protect\citeauthoryear{{Frank}, {King}  \& {Raine}}{{Frank}
  et~al.}{2002}]{Frank+2002}
{Frank} J.,  {King} A.,   {Raine} D.~J.,  2002, {Accretion Power in
  Astrophysics: Third Edition}

\bibitem[\protect\citeauthoryear{{H{\= o}shi}}{{H{\= o}shi}}{1973}]{Hoshi1973}
{H{\= o}shi} R.,  1973, \mn@doi [Progress of Theoretical Physics]
  {10.1143/PTP.49.776}, \href
  {http://adsabs.harvard.edu/abs/1973PThPh..49..776H} {49, 776}

\bibitem[\protect\citeauthoryear{{Hameury}, {Bonazzola}, {Heyvaerts}  \&
  {Lasota}}{{Hameury} et~al.}{1983}]{Hameury+1983}
{Hameury} J.~M.,  {Bonazzola} S.,  {Heyvaerts} J.,   {Lasota} J.~P.,  1983,
  \aap, \href {http://adsabs.harvard.edu/abs/1983A%26A...128..369H} {128, 369}

\bibitem[\protect\citeauthoryear{{Howell}, {Walter}, {Harrison}, {Huber},
  {Becker}  \& {White}}{{Howell} et~al.}{2006}]{Howell+2006b}
{Howell} S.~B.,  {Walter} F.~M.,  {Harrison} T.~E.,  {Huber} M.~E.,  {Becker}
  R.~H.,   {White} R.~L.,  2006, \mn@doi [\apj] {10.1086/507603}, \href
  {http://adsabs.harvard.edu/abs/2006ApJ...652..709H} {652, 709}

\bibitem[\protect\citeauthoryear{{Imamura}}{{Imamura}}{1985}]{Imamura1985}
{Imamura} J.~N.,  1985, \mn@doi [\apj] {10.1086/163426}, \href
  {http://adsabs.harvard.edu/abs/1985ApJ...296..128I} {296, 128}

\bibitem[\protect\citeauthoryear{{Imamura}, {Wolff}  \& {Durisen}}{{Imamura}
  et~al.}{1984}]{Imamura+1984}
{Imamura} J.~N.,  {Wolff} M.~T.,   {Durisen} R.~H.,  1984, \mn@doi [\apj]
  {10.1086/161654}, \href {http://adsabs.harvard.edu/abs/1984ApJ...276..667I}
  {276, 667}

\bibitem[\protect\citeauthoryear{{Imamura}, {Durisen}, {Lamb}  \&
  {Weast}}{{Imamura} et~al.}{1987}]{Imamura+1987}
{Imamura} J.~N.,  {Durisen} R.~H.,  {Lamb} D.~Q.,   {Weast} G.~J.,  1987,
  \mn@doi [\apj] {10.1086/164969}, \href
  {http://adsabs.harvard.edu/abs/1987ApJ...313..298I} {313, 298}

\bibitem[\protect\citeauthoryear{{Imamura}, {Rashed}  \& {Wolff}}{{Imamura}
  et~al.}{1991}]{Imamura+1991}
{Imamura} J.~N.,  {Rashed} H.,   {Wolff} M.~T.,  1991, \mn@doi [\apj]
  {10.1086/170466}, \href {http://adsabs.harvard.edu/abs/1991ApJ...378..665I}
  {378, 665}

\bibitem[\protect\citeauthoryear{{Imamura}, {Aboasha}, {Wolff}  \&
  {Wood}}{{Imamura} et~al.}{1996}]{Imamura+1996}
{Imamura} J.~N.,  {Aboasha} A.,  {Wolff} M.~T.,   {Wood} K.~S.,  1996, \mn@doi
  [\apj] {10.1086/176815}, \href
  {http://adsabs.harvard.edu/abs/1996ApJ...458..327I} {458, 327}

\bibitem[\protect\citeauthoryear{{Imamura}, {Steiman-Cameron}  \&
  {Wolff}}{{Imamura} et~al.}{2000}]{Imamura+2000}
{Imamura} J.~N.,  {Steiman-Cameron} T.~Y.,   {Wolff} M.~T.,  2000, \mn@doi
  [\pasp] {10.1086/316478}, \href
  {http://adsabs.harvard.edu/abs/2000PASP..112...18I} {112, 18}

\bibitem[\protect\citeauthoryear{{King} \& {Lasota}}{{King} \&
  {Lasota}}{1979}]{King+Lasota1979}
{King} A.~R.,  {Lasota} J.~P.,  1979, \mn@doi [\mnras]
  {10.1093/mnras/188.3.653}, \href
  {http://adsabs.harvard.edu/abs/1979MNRAS.188..653K} {188, 653}

\bibitem[\protect\citeauthoryear{{Lamb} \& {Masters}}{{Lamb} \&
  {Masters}}{1979}]{Lamb+Masters1979}
{Lamb} D.~Q.,  {Masters} A.~R.,  1979, \mn@doi [\apjl] {10.1086/183121}, \href
  {http://adsabs.harvard.edu/abs/1979ApJ...234L.117L} {234, L117}

\bibitem[\protect\citeauthoryear{{Landi}, {Bassani}, {Dean}, {Bird}, {Fiocchi},
  {Bazzano}, {Nousek}  \& {Osborne}}{{Landi} et~al.}{2009}]{Landi+2009}
{Landi} R.,  {Bassani} L.,  {Dean} A.~J.,  {Bird} A.~J.,  {Fiocchi} M.,
  {Bazzano} A.,  {Nousek} J.~A.,   {Osborne} J.~P.,  2009, \mn@doi [\mnras]
  {10.1111/j.1365-2966.2008.14086.x}, \href
  {http://adsabs.harvard.edu/abs/2009MNRAS.392..630L} {392, 630}

\bibitem[\protect\citeauthoryear{{Langer}, {Chanmugam}  \& {Shaviv}}{{Langer}
  et~al.}{1981}]{Langer+1981}
{Langer} S.~H.,  {Chanmugam} G.,   {Shaviv} G.,  1981, \mn@doi [\apjl]
  {10.1086/183514}, \href {http://adsabs.harvard.edu/abs/1981ApJ...245L..23L}
  {245, L23}

\bibitem[\protect\citeauthoryear{{Langer}, {Chanmugam}  \& {Shaviv}}{{Langer}
  et~al.}{1982}]{Langer+1982}
{Langer} S.~H.,  {Chanmugam} C.,   {Shaviv} G.,  1982, \mn@doi [\apj]
  {10.1086/160079}, \href {http://adsabs.harvard.edu/abs/1982ApJ...258..289L}
  {258, 289}

\bibitem[\protect\citeauthoryear{{Larsson}}{{Larsson}}{1985}]{Larsson1985}
{Larsson} S.,  1985, \aap, \href
  {http://adsabs.harvard.edu/abs/1985A%26A...145L...1L} {145, L1}

\bibitem[\protect\citeauthoryear{{Larsson}}{{Larsson}}{1987}]{Larsson1987}
{Larsson} S.,  1987, \aap, \href
  {http://adsabs.harvard.edu/abs/1987A%26A...181L..15L} {181, L15}

\bibitem[\protect\citeauthoryear{{Larsson}}{{Larsson}}{1989}]{Larsson1989}
{Larsson} S.,  1989, \aap, \href
  {http://adsabs.harvard.edu/abs/1989A%26A...217..146L} {217, 146}

\bibitem[\protect\citeauthoryear{{Litwin}, {Brown}  \& {Rosner}}{{Litwin}
  et~al.}{2001}]{Litwin+2001}
{Litwin} C.,  {Brown} E.~F.,   {Rosner} R.,  2001, \mn@doi [\apj]
  {10.1086/320952}, \href {http://adsabs.harvard.edu/abs/2001ApJ...553..788L}
  {553, 788}

\bibitem[\protect\citeauthoryear{{Longair}}{{Longair}}{2011}]{Longair2011}
{Longair} M.~S.,  2011, {High Energy Astrophysics}

\bibitem[\protect\citeauthoryear{{Mason}, {Jensen}, {Murdin}, {Clark},
  {Middleditch}, {Cordova}, {Reichert}  \& {Bowyer}}{{Mason}
  et~al.}{1983}]{Mason+1983}
{Mason} K.~O.,  {Jensen} K.~A.,  {Murdin} P.~G.,  {Clark} D.,  {Middleditch}
  J.,  {Cordova} F.~A.,  {Reichert} G.,   {Bowyer} S.,  1983, \mn@doi [\apj]
  {10.1086/160627}, \href {http://adsabs.harvard.edu/abs/1983ApJ...264..575M}
  {264, 575}

\bibitem[\protect\citeauthoryear{{Masters}, {Fabian}, {Rees}  \&
  {Pringle}}{{Masters} et~al.}{1977}]{Masters+1977}
{Masters} A.~R.,  {Fabian} A.~C.,  {Rees} M.~J.,   {Pringle} J.~E.,  1977,
  \mn@doi [\mnras] {10.1093/mnras/178.3.501}, \href
  {http://adsabs.harvard.edu/abs/1977MNRAS.178..501M} {178, 501}

\bibitem[\protect\citeauthoryear{{Middleditch}}{{Middleditch}}{1982}]{Middleditch1982}
{Middleditch} J.,  1982, \mn@doi [\apjl] {10.1086/183811}, \href
  {http://adsabs.harvard.edu/abs/1982ApJ...257L..71M} {257, L71}

\bibitem[\protect\citeauthoryear{{Middleditch}, {Imamura}  \&
  {Steiman-Cameron}}{{Middleditch} et~al.}{1997}]{Middleditch+1997}
{Middleditch} J.,  {Imamura} J.~N.,   {Steiman-Cameron} T.~Y.,  1997, \apj,
  \href {http://adsabs.harvard.edu/abs/1997ApJ...489..912M} {489, 912}

\bibitem[\protect\citeauthoryear{{Mignone}}{{Mignone}}{2005}]{Mignone2005}
{Mignone} A.,  2005, \mn@doi [\apj] {10.1086/429905}, \href
  {http://adsabs.harvard.edu/abs/2005ApJ...626..373M} {626, 373}

\bibitem[\protect\citeauthoryear{{Mignone}, {Bodo}, {Massaglia}, {Matsakos},
  {Tesileanu}, {Zanni}  \& {Ferrari}}{{Mignone} et~al.}{2007}]{Mignone+2007}
{Mignone} A.,  {Bodo} G.,  {Massaglia} S.,  {Matsakos} T.,  {Tesileanu} O.,
  {Zanni} C.,   {Ferrari} A.,  2007, \mn@doi [\apjs] {10.1086/513316}, \href
  {http://adsabs.harvard.edu/abs/2007ApJS..170..228M} {170, 228}

\bibitem[\protect\citeauthoryear{{Mouchet} et~al.,}{{Mouchet}
  et~al.}{2017}]{Mouchet+2017}
{Mouchet} M.,  et~al., 2017, \mn@doi [\aap] {10.1051/0004-6361/201630166},
  \href {http://adsabs.harvard.edu/abs/2017A%26A...600A..53M} {600, A53}

\bibitem[\protect\citeauthoryear{{Mukherjee} \& {Bhattacharya}}{{Mukherjee} \&
  {Bhattacharya}}{2012}]{Mukherjee+2012}
{Mukherjee} D.,  {Bhattacharya} D.,  2012, \mn@doi [\mnras]
  {10.1111/j.1365-2966.2011.20085.x}, \href
  {http://adsabs.harvard.edu/abs/2012MNRAS.420..720M} {420, 720}

\bibitem[\protect\citeauthoryear{{Mukherjee}, {Bhattacharya}  \&
  {Mignone}}{{Mukherjee} et~al.}{2013a}]{Mukherjee+2013a}
{Mukherjee} D.,  {Bhattacharya} D.,   {Mignone} A.,  2013a, \mn@doi [\mnras]
  {10.1093/mnras/stt020}, \href
  {http://adsabs.harvard.edu/abs/2013MNRAS.430.1976M} {430, 1976}

\bibitem[\protect\citeauthoryear{{Mukherjee}, {Bhattacharya}  \&
  {Mignone}}{{Mukherjee} et~al.}{2013b}]{Mukherjee+2013b}
{Mukherjee} D.,  {Bhattacharya} D.,   {Mignone} A.,  2013b, \mn@doi [\mnras]
  {10.1093/mnras/stt1344}, \href
  {http://adsabs.harvard.edu/abs/2013MNRAS.435..718M} {435, 718}

\bibitem[\protect\citeauthoryear{{Payne} \& {Melatos}}{{Payne} \&
  {Melatos}}{2004}]{Payne+Melatos2004}
{Payne} D.~J.~B.,  {Melatos} A.,  2004, \mn@doi [\mnras]
  {10.1111/j.1365-2966.2004.07798.x}, \href
  {http://adsabs.harvard.edu/abs/2004MNRAS.351..569P} {351, 569}

\bibitem[\protect\citeauthoryear{{Payne} \& {Melatos}}{{Payne} \&
  {Melatos}}{2007}]{Payne+Melatos2007}
{Payne} D.~J.~B.,  {Melatos} A.,  2007, \mn@doi [\mnras]
  {10.1111/j.1365-2966.2007.11451.x}, \href
  {http://adsabs.harvard.edu/abs/2007MNRAS.376..609P} {376, 609}

\bibitem[\protect\citeauthoryear{{Potter} et~al.,}{{Potter}
  et~al.}{2010}]{Potter+2010}
{Potter} S.~B.,  et~al., 2010, \mn@doi [\mnras]
  {10.1111/j.1365-2966.2009.15944.x}, \href
  {http://adsabs.harvard.edu/abs/2010MNRAS.402.1161P} {402, 1161}

\bibitem[\protect\citeauthoryear{{Press}, {Teukolsky}, {Vetterling}  \&
  {Flannery}}{{Press} et~al.}{2007}]{Press+2007}
{Press} W.~H.,  {Teukolsky} S.~A.,  {Vetterling} W.~T.,   {Flannery} B.~P.,
  2007, Numerical Recipes 3rd Edition: The Art of Scientific Computing, 3 edn.
Cambridge University Press, New York, NY, USA

\bibitem[\protect\citeauthoryear{{Rybicki} \& {Lightman}}{{Rybicki} \&
  {Lightman}}{1979}]{Rybicki+Lightman1979}
{Rybicki} G.~B.,  {Lightman} A.~P.,  1979, {Radiative processes in
  astrophysics}

\bibitem[\protect\citeauthoryear{{Saxton} \& {Wu}}{{Saxton} \&
  {Wu}}{1999}]{Saxton+Wu1999}
{Saxton} C.~J.,  {Wu} K.,  1999, \mn@doi [\mnras]
  {10.1046/j.1365-8711.1999.02967.x}, \href
  {http://adsabs.harvard.edu/abs/1999MNRAS.310..677S} {310, 677}

\bibitem[\protect\citeauthoryear{{Saxton}, {Wu}, {Pongracic}  \&
  {Shaviv}}{{Saxton} et~al.}{1998}]{Saxton+1998}
{Saxton} C.~J.,  {Wu} K.,  {Pongracic} H.,   {Shaviv} G.,  1998, \mn@doi
  [\mnras] {10.1046/j.1365-8711.1998.01839.x}, \href
  {http://adsabs.harvard.edu/abs/1998MNRAS.299..862S} {299, 862}

\bibitem[\protect\citeauthoryear{{Saxton}, {Wu}, {Cropper}  \&
  {Ramsay}}{{Saxton} et~al.}{2005}]{Saxton+2005}
{Saxton} C.~J.,  {Wu} K.,  {Cropper} M.,   {Ramsay} G.,  2005, \mn@doi [\mnras]
  {10.1111/j.1365-2966.2005.09103.x}, \href
  {http://adsabs.harvard.edu/abs/2005MNRAS.360.1091S} {360, 1091}

\bibitem[\protect\citeauthoryear{{Schwarz} et~al.,}{{Schwarz}
  et~al.}{1998}]{Schwarz+1998}
{Schwarz} R.,  et~al., 1998, \aap, \href
  {http://adsabs.harvard.edu/abs/1998A%26A...338..465S} {338, 465}

\bibitem[\protect\citeauthoryear{{Schwope} \& {Beuermann}}{{Schwope} \&
  {Beuermann}}{1990}]{Schwope+Beuermann1990}
{Schwope} A.~D.,  {Beuermann} K.,  1990, \aap, \href
  {http://adsabs.harvard.edu/abs/1990A%26A...238..173S} {238, 173}

\bibitem[\protect\citeauthoryear{{Schwope}, {Thomas}, {Beuermann}  \&
  {Reinsch}}{{Schwope} et~al.}{1993}]{Schwope+1993}
{Schwope} A.~D.,  {Thomas} H.-C.,  {Beuermann} K.,   {Reinsch} K.,  1993, \aap,
  \href {http://adsabs.harvard.edu/abs/1993A%26A...267..103S} {267, 103}

\bibitem[\protect\citeauthoryear{{Spitzer}}{{Spitzer}}{1962}]{Spitzer1962}
{Spitzer} L.,  1962, {Physics of Fully Ionized Gases}

\bibitem[\protect\citeauthoryear{{Strickland} \& {Blondin}}{{Strickland} \&
  {Blondin}}{1995}]{Strickland+Blondin1995}
{Strickland} R.,  {Blondin} J.~M.,  1995, \mn@doi [\apj] {10.1086/176093},
  \href {http://adsabs.harvard.edu/abs/1995ApJ...449..727S} {449, 727}

\bibitem[\protect\citeauthoryear{{Sutherland}, {Bicknell}  \&
  {Dopita}}{{Sutherland} et~al.}{2003}]{Sutherland+2003}
{Sutherland} R.~S.,  {Bicknell} G.~V.,   {Dopita} M.~A.,  2003, \mn@doi [\apj]
  {10.1086/375294}, \href {http://adsabs.harvard.edu/abs/2003ApJ...591..238S}
  {591, 238}

\bibitem[\protect\citeauthoryear{{Visvanathan} \&
  {Wickramasinghe}}{{Visvanathan} \&
  {Wickramasinghe}}{1979}]{Visvanathan+Wickramasinghe1979}
{Visvanathan} N.,  {Wickramasinghe} D.~T.,  1979, \mn@doi [\nat]
  {10.1038/281047a0}, \href {http://adsabs.harvard.edu/abs/1979Natur.281...47V}
  {281, 47}

\bibitem[\protect\citeauthoryear{{Wada}, {Shimizu}, {Suzuki}, {Kato}  \&
  {Hoshi}}{{Wada} et~al.}{1980}]{Wada+1980}
{Wada} T.,  {Shimizu} A.,  {Suzuki} M.,  {Kato} M.,   {Hoshi} R.,  1980,
  \mn@doi [Progress of Theoretical Physics] {10.1143/PTP.64.1986}, \href
  {http://adsabs.harvard.edu/abs/1980PThPh..64.1986W} {64, 1986}

\bibitem[\protect\citeauthoryear{{Wolff}, {Imamura}, {Middleditch}, {Wood}  \&
  {Steiman-Cameron}}{{Wolff} et~al.}{1999}]{Wolff+1999}
{Wolff} M.~T.,  {Imamura} J.~N.,  {Middleditch} J.,  {Wood} K.~S.,
  {Steiman-Cameron} T.,  1999, in {Hellier} C.,  {Mukai} K.,  eds,
  Astronomical Society of the Pacific Conference Series Vol. 157, Annapolis
  Workshop on Magnetic Cataclysmic Variables. p.~149

\bibitem[\protect\citeauthoryear{{Worpel} \& {Schwope}}{{Worpel} \&
  {Schwope}}{2015}]{Worpel+Schwope2015}
{Worpel} H.,  {Schwope} A.~D.,  2015, \mn@doi [\aap]
  {10.1051/0004-6361/201526916}, \href
  {http://adsabs.harvard.edu/abs/2015A%26A...583A.130W} {583, A130}

\bibitem[\protect\citeauthoryear{{Wu}, {Chanmugam}  \& {Shaviv}}{{Wu}
  et~al.}{1992}]{Wu+1992}
{Wu} K.,  {Chanmugam} G.,   {Shaviv} G.,  1992, \mn@doi [\apj]
  {10.1086/171782}, \href {http://adsabs.harvard.edu/abs/1992ApJ...397..232W}
  {397, 232}

\bibitem[\protect\citeauthoryear{{Wu}, {Chanmugam}  \& {Shaviv}}{{Wu}
  et~al.}{1994}]{Wu+1994}
{Wu} K.,  {Chanmugam} G.,   {Shaviv} G.,  1994, \mn@doi [\apj]
  {10.1086/174103}, \href {http://adsabs.harvard.edu/abs/1994ApJ...426..664W}
  {426, 664}

\bibitem[\protect\citeauthoryear{{Wu}, {Chanmugam}  \& {Shaviv}}{{Wu}
  et~al.}{1995}]{Wu+1995}
{Wu} K.,  {Chanmugam} G.,   {Shaviv} G.,  1995, \mn@doi [\apj]
  {10.1086/176574}, \href {http://adsabs.harvard.edu/abs/1995ApJ...455..260W}
  {455, 260}

\bibitem[\protect\citeauthoryear{{de Martino}, {Matt}, {Belloni}, {Haberl}  \&
  {Mukai}}{{de Martino} et~al.}{2004}]{deMartino+2004}
{de Martino} D.,  {Matt} G.,  {Belloni} T.,  {Haberl} F.,   {Mukai} K.,  2004,
  \mn@doi [\aap] {10.1051/0004-6361:20034160}, \href
  {http://adsabs.harvard.edu/abs/2004A%26A...415.1009D} {415, 1009}

\makeatother
\end{thebibliography}
\label{lastpage}
\end{document}